\documentclass[showpacs,amssymb,prc,preprint]{revtex4}
\usepackage[dvips]{graphicx}

\begin{document}
\title{Relating pseudospin and spin symmetries through charge conjugation and chiral transformations:
the case of the relativistic harmonic oscillator}
\author{A. S. de Castro}
\affiliation{Departamento de F\'{\i}sica e Qu\'{\i}mica, Universidade Estadual Paulista,
 12516-410 Guaratinguet\'a, SP, Brazil\\ and
Departamento de F\'{\i}sica and Centro de F\'{\i}sica
Computacional, Universidade de Coimbra, P-3004-516 Coimbra,
Portugal}
\author{P. Alberto, R. Lisboa}
\affiliation{Departamento de F\'{\i}sica and Centro de F\'{\i}sica
Computacional, Universidade de Coimbra, P-3004-516 Coimbra,
Portugal}
\author{M. Malheiro}
\affiliation{Departamento de F\'{\i}sica,
Instituto Tecnol\'ogico de Aeron\'autica, CTA,
12228-900, S\~ao Jos\'e dos Campos, SP, Brazil \\ and
Instituto de F\'{\i}sica, Universidade Federal
Fluminense, 24210-340 Niter\'oi, Brazil}
\date{\today}


\begin{abstract}
\noindent We solve the generalized relativistic harmonic oscillator in 1+1
dimensions, i.e., including a linear pseudoscalar potential and quadratic
scalar and vector potentials which have equal or opposite signs. We consider positive and
negative quadratic potentials and discuss in detail their bound-state solutions for fermions
and antifermions. The main features of these bound states are the
same as the ones of the generalized three-dimensional
relativistic harmonic oscillator bound states. The solutions found for
zero pseudoscalar potential
are related to the spin and pseudospin symmetry of the Dirac equation in
3+1 dimensions. We show how the charge conjugation and $\gamma^5$ chiral
transformations relate the several spectra obtained and find that
for massless particles the spin and pseudospin symmetry related problems have the same spectrum,
but different spinor solutions.
Finally, we establish a relation of the solutions found with single-particle states of nuclei
described by relativistic mean-field theories with scalar, vector and isoscalar tensor
interactions and discuss the conditions
in which one may have both nucleon and antinucleon bound states.

\end{abstract}
\maketitle

\section{Introduction}

Recently, there has been a wide interest in relativistic potentials
involving mixtures of vector and scalar potentials with opposite signs. The
interest lies on attempts to explain the pseudospin symmetry in nuclear
physics. Chen \textit{et al.} \cite{che}, using a Dirac Hamiltonian with
scalar $V_{s}$ and vector $V_{v}$ potentials quadratic in space coordinates,
found a harmonic-oscillator-like second order equation which can be solved
analytically for $\Delta =V_{t}-V_{s}=0$, as considered before by Kukulin
\textit{et al.} \cite{kuk}, and also for $\Sigma =V_{t}+V_{s}=0$. Very
recently, Ginocchio solved the triaxial, axial, and spherical harmonic
oscillators for the case $\Delta =0$ and applied it to the study of
antinucleons embedded in nuclei \cite{gin2}.
The case $\Sigma=0$ is particularly relevant in nuclear physics,
since it is usually pointed out as
a necessary condition for occurrence of pseudospin symmetry in nuclei \cite
{gin3}-\cite{gin5}. Also the observed isospin dependence of pseudospin can
be explained by the effect of the $\rho$ potential on the shape of
the $\Sigma$ potential \cite{meng0}-\cite{bjp34}. As shown by Bell and Ruegg \cite{bell},
the $\Delta=0$ and $\Sigma=0$ cases correspond to a $SU(2)$
symmetry of the Dirac Hamiltonian with only scalar and vector potentials which is
independent of the particular shapes of these
potentials.

A generalized harmonic oscillator (HO) for spin 1/2 particles that
includes not only those combinations of scalar and vector potentials,
but also a linear tensor potential, giving rise to the so-called
Dirac oscillator, has also been considered in \cite{nosso}.
The research of this kind of interaction was started
by It\^o \cite{ito} and has been revived by Moshinsky and Szczepaniak
\cite{mosh1} (see \cite{nosso} for a complete reference list).

In Ref.~\cite{nosso} a special attention was paid to the case
when the scalar and vector potentials are such that $\Sigma =0$, which,
as stated before, has been related to the
nuclear pseudospin symmetry. There it has been concluded that only
when the tensor potential is turned on one gets negative energy
bound-state solutions along with the usual positive energy
solutions. Those negative energy solutions are important to describe the
possible existence of
antinucleons in nuclei \cite{gin5}. When the tensor potential is absent the
special conditions relating the scalar and vector potentials
($V_{s}=\pm V_{t}$) are needed to have a standard
positive HO potential which is only able to bind fermions,
and therefore exclude the negative
bound-state solutions from the spectra. Actually, as was discussed in another
context in \cite{asc3}, with an appropriate choice of signs, potentials fulfilling
the relations $V_{s}=\pm V_{t}$ are able to bind either fermions or antifermions.
As an example, we will see later that a (negative)
harmonic potential, with no tensor potential, binds antifermions and does not
allow positive bound-state solutions.

Closely related to this is the fact that, in the nucleus, the
charge-conjugation transformation relates the spin symmetry of the negative
bound-state solutions (antinucleons) to the pseudospin symmetry of the
positive bound-state solutions (nucleons) \cite{Zhou}. This has also been
discussed recently in \cite{gin2}, analyzing the HO for
antinucleons with spin symmetry ($\Delta =0$) and without
the tensor interaction. Again, the charge conjugation connects this problem
with the HO for nucleons with pseudospin symmetry ($\Sigma
=0$), but in this case the positive harmonic binding potential is $\Delta$.
Note that, because the harmonic potential does not vanish for large
distances but is a confining potential, it is possible to obtain particle
bound-state solutions in the exact pseudospin limit \cite{nosso}.
Therefore, we believe that this connection between
spin and pseudospin symmetry obtained by charge conjugation deserves to
be more explored. In addition, after the pseudospin symmetry has been
revealed as a relativistic symmetry by Ginocchio \cite{gin3}, a link of it to chiral
symmetry has been suggested based on sum rules calculations in relativistic
chiral models that support a small value for $\Sigma$ \cite{furn}.
However, until now, a clear connection of pseudospin and chiral symmetries
has not been established.

The main motivation of this paper is to shed some light in the relation between
spin and pseudospin symmetries by means of charge-conjugation and $\gamma^5$
chiral transformations.
In order to do so, we solve the mixed scalar-vector-pseudoscalar HO potential
in 1+1 dimensions, taking advantage of the simplicity
of the lowest dimensionality of the space-time. This approach is equivalent
to consider fermions in 3+1 dimensions which are restricted to move in one direction
\cite{str}. We explore the spectra when
it is possible to obtain analytical solutions, \textit{i.e.}, in the
particular cases when $\Delta =0$ and $\Sigma=0$. As referred before, if the
pseudocalar term is turned off, these cases correspond, respectively, to have exact spin
and pseudospin symmetry in 3+1 dimensions. We explore all the possible signs of the
quadratic ($\Delta $ or $\Sigma $) and linear (pseudoscalar) potentials,
thus paying attention to bound states of fermions and antifermions as well.
We compare both cases $\Delta=0$ and $\Sigma=0$ to establish the
charge-conjugation connection discussed above in the presence of the
pseudoscalar term. We also consider the case of zero mass to look for the
connection between spin and pseudospin symmetries by means of the chiral
transformation.

An added motivation for this paper is given by the recent demonstration that the
relativistic 3+1 HO with scalar and vector potentials
can describe successfully the heavy nucleus spectrum \cite{guo}.
The parameters of the HO are determined by fitting the scalar and vector potentials derived
from Relativistic Mean Field nuclear calculations (RMF).
The $^{208}$Pb neutron single-particle levels obtained in the HO are very similar
to the RMF results. The pseudospin symmetry is shown to be almost satisfied for a heavy nucleus
in the HO description. The possibility of breaking perturbatively this symmetry in the 3+1 HO
by a tensor interaction (in our 1+1 case a pseudoscalar coupling) has been discussed in
\cite{ronai} and can be included in a calculation like the one of Ref.~\cite{guo}
in order to improve the results. Thus, the study of all the possible eigenenergies
 of the 1+1 HO, presented in this paper, considering not only the positive energy
 solutions already obtained for the 3+1 case in \cite{nosso} but also the negative ones,
 can be applied to analyze the existence of antinucleons in nuclei and their spectrum.

The particle and antiparticle spectra we obtain for the cases $\Sigma=0$ and
$\Delta=0$ demonstrate the charge-conjugation connection referred above.
They also show that the spectra of massless $\Delta =0$ states are
degenerate with massless $\Sigma =0$ states. Further imposing chiral
symmetry to these states, such that they would become eigenstates of $\gamma^5$,
would imply that all potentials would be zero and
thus one would have just massless free fermions. Our conclusions are valid in
the case of 3+1 dimensions, because, as will be shown later,
one basically has to add the angular momentum quantum numbers $\ell$ and $\kappa$
to the coefficients of the functions appearing in the 1+1 eigenvalue equations in order to
get the corresponding 3+1 equations.

This paper is organized as follows. In Sec.~\ref{Sec:Dirac_1+1} we
present the general Dirac equation in 1+1 dimensions with a potential with a
completely general Lorentz structure. The effect of charge-conjugation
and the $\gamma^5$ chiral transformations upon the general 1+1 Dirac Hamiltonian
is discussed in Secs.~\ref{SubSec:charge_conj} and \ref{SubSec:chiral}
respectively. In Secs.~\ref{SubSec:Eq_motion} and \ref{SubSec:Sturm-Liouville}
we present the equations of motion and discuss the isolated solutions, {\it i.e.},
the solutions out of the Sturm-Liouville problem, for $\Delta=0$ and $\Sigma=0$.
The nonrelativistic limits of the
$\Delta=0$ and $\Sigma=0$ cases are discussed in Sec.~\ref{SubSec:nonrel_limit},
where we show that when $\Sigma=0$ there is no nonrelativistic limit when the
pseudoscalar potential vanishes or is very small.
In Sec.~\ref{Sec:rel_h_oscillator} we introduce the general relativistic
harmonic oscillator in 1+1 dimensions for $\Delta=0$ by assigning a quadratic potential
$k_1x^2/2$ for $\Sigma$ and a linear potential $k_2 x$ for the pseudoscalar potential $V_p$.
There we also present the eigenvalue equation and the wave functions for bound states,
discuss the Dirac oscillator case ($\Delta=\Sigma=0$ and $V_p\not=0$) and show how one
can use the charge-conjugation and chiral transformations to get the eigenenergies
for the $\Sigma=0$ case from the $\Delta=0$ case.
The following section is devoted to a detailed analysis of the solutions of the $\Delta=0$
relativistic HO for all signs of the parameters $k_1$ and $k_2$.
Finally in Sec.~\ref{Sec:sol_sym_arg} we discuss the spectrum for $\Sigma=0$ and
massless fermions and in Sec.~\ref{Sec:conclusions} we draw the conclusions.

\section{The Dirac equation in 1+1 dimensions}

\label{Sec:Dirac_1+1}

The two-dimensional Dirac equation can be obtained from the
four-dimen\-sional one with the mixture of spherically symmetric scalar,
vector and anomalous magnetic-like (tensor) interactions. If we limit the
fermion motion to the $x$-direction ($p_{y}=p_{z}=0$) the four-dimensional
Dirac equation decomposes into two equivalent two-dimensional equations with
2-component spinors and 2$\times $2 matrices \cite{str}. Then, there results
that the scalar and vector interactions preserve their Lorentz structures
whereas the anomalous magnetic-like interaction turns out to be a
pseudoscalar interaction. Furthermore, in the 1+1 world there is no angular
momentum so that the spin is absent. Therefore, the 1+1 dimensional Dirac
equation allows us to explore the physical consequences of the
negative-energy states in a mathematically simpler and more physically
transparent way. In this spirit the two-dimensional version of the anomalous
magnetic-like interaction linear in the space coordinate has also received
attention \cite{nogami}-\cite{asc1}. Later this system was shown to be a
particular case of a more general class of exactly solvable problems \cite%
{asc2}.

Let us begin by presenting the Dirac equation in 1+1 dimensions. The Dirac
equation for a particle with mass $m$ and a general potential $\mathcal{V}$,
to be defined later, is formally the same as its 3+1 counterpart:
\begin{equation}
i\hbar c\gamma ^{\mu }\partial _{\mu }\Psi -mc^{2}\Psi -\gamma ^{0}\mathcal{V%
}\Psi =0\ .  \label{eq_Dirac_gen}
\end{equation}%
Here $\mu =0,1$ and the $2\times 2$ matrices $\gamma ^{\mu }$ obey the usual
relations $\{\gamma ^{\mu },\gamma ^{\nu }\}=2g^{\mu \nu }I$, where $I$ is
the $2\times 2$ identity matrix. The metric tensor and the covariant
derivative operator are, respectively, $g^{\mu \nu }=\mathrm{{diag}(1,-1)}$
and $\partial _{\mu }=\big(\partial /\partial (ct)\,,\,\partial /\partial x%
\big)$. The Hamiltonian is defined in the usual way, such that Eq. (\ref%
{eq_Dirac_gen}) is written as
\begin{equation}
i\hbar \frac{\partial \Psi }{\partial t}=H\Psi =\big(c\alpha \,p+\beta
mc^{2}+\mathcal{V}\big)\Psi \ ,  \label{def_hamiltoniano_d}
\end{equation}%
where $p\equiv -i\hbar \,\partial /\partial x$. The positive definite
function $|\Psi |^{2}=\Psi ^{\dagger }\Psi $, satisfying a continuity
equation, is interpreted as a position probability density and its norm is a
constant of motion. This interpretation is completely satisfactory for
single-particle states \cite{tha}. The traceless matrices $\alpha $ and $%
\beta $ are defined by $\alpha =\gamma ^{0}\gamma ^{1}$ and $\beta =\gamma
^{0}$, and obey the relations $\alpha ^{2}=\beta ^{2}=I$, $\{\alpha ,\beta
\}=0$. We set $\mathcal{V}$ to be
\begin{equation}
\mathcal{V}=I\,V_{t}+\beta V_{s}+\alpha V_{sp}-i\beta \gamma ^{5}V_{p}\ ,
\label{def_V}
\end{equation}%
where $\gamma ^{5}=\gamma ^{0}\gamma ^{1}=\alpha $. This is the most general
combination of Lorentz structures because there are only four linearly
independent 2$\times $2 matrices. The subscripts for the terms of the
potential denote their properties under a Lorentz transformation: $t$ and $%
sp $ for the time and space components of the 2-vector potential, $s$ and $p$
for the scalar and pseudoscalar terms, respectively.

If the terms in the potential $\mathcal{V}$ are time-independent, the Dirac
equation (\ref{def_hamiltoniano_d}) becomes
\begin{equation}
H\tilde\psi =E\tilde\psi  \label{eq1}
\end{equation}
where
\begin{eqnarray}
H&=&c\alpha p+\beta mc^{2}+I\,V_{t}(x)+\alpha V_{sp}(x)+\beta V_{s}(x)-i\beta
\gamma ^{5}V_{p}(x)  \label{time_ind_H} \ ,\\
\tilde\psi(x)&=&e^{\frac{i}{\hbar}E t}\Psi(x,t)  \label{time_ind_psi}
\end{eqnarray}
and $E$ is the fermion energy. To have an explicit expression for the
$\alpha$ and $\beta$ matrices one can choose 2$\times $2 Pauli matrices
which satisfy the same algebra. We use $\beta=\sigma_3$, $\alpha=\sigma_1$
and thus $\beta\gamma^5=i\sigma_2$.

The Hamiltonian (\ref{time_ind_H}) is invariant under the parity operation,
\textit{i.e.}, when $x\rightarrow -x$, $V_{sp}(x)$ and $V_{p}(x)$ change
sign, whereas $V_{t}(x)$ and $V_{s}(x)$ remain the same. This is because the
parity operator is $P=\exp (i\eta )P_{0}\sigma_{3}$, where $\eta$ is a
constant phase and $P_{0}$ changes $x$ into $-x$. Since this unitary
operator anticommutes with $\alpha$ and $\beta \gamma ^{5}$, they change
sign under a parity transformation, whereas $I$ and $\beta$, which commute
with $P$, remain the same. When one writes down the explicit equations of
motion in terms of the components of the spinor $\tilde\psi$, the
combinations $\Sigma=V_t+V_s$ and $\Delta=V_t-V_s$ of the vector and scalar
components arise naturally. Therefore, it is convenient to rewrite the
Hamiltonian (\ref{time_ind_H}) in terms of these potentials. We have
\begin{equation}  \label{hamiltonian_d_sigma_delta}
H = c\alpha\,p+\beta mc^2 +\alpha V_{sp}+\frac{I+\beta}2\Sigma+ \frac{I-\beta%
}2\Delta-i\beta\gamma^5 V_p\ .
\end{equation}

\subsection{Charge conjugation}

\label{SubSec:charge_conj}

The charge-conjugation operation changes the sign of the electromagnetic
interaction, \textit{i.e.}, changes the sign of the time and space
potentials in (\ref{def_V}). This is accomplished by the transformation
(see, for example, Itzykson and Zuber \cite{Itzykson_Zuber})
\begin{equation}
\Psi \ \longrightarrow \ \Psi _{c}=C\bar{\Psi}^{T}=C{\gamma ^{0}\,}^{T}\Psi
^{\ast }\ ,  \label{conj_carga}
\end{equation}%
where $T$ denotes matrix transposition and $C$ is a matrix such that $C{%
\gamma ^{\mu }\,}^{T}C^{-1}=-\gamma ^{\mu }$. In $1+1$ dimensions one matrix
that satisfies this relation is
\[
C=e^{i\theta }\alpha \beta \ .
\]%
If we choose the phase factor $e^{i\theta }$ equal to 1, we have
\begin{equation}
\Psi _{c}=C\bar{\Psi}^{T}=\alpha \Psi ^{\ast } \ .
\end{equation}%
After applying this charge-conjugation operation to the Dirac equation (\ref%
{def_hamiltoniano_d}), the time-independent Dirac equation becomes
\begin{equation}
H_{c}\tilde{\psi}_{c}=-E\tilde{\psi}_{c}\ ,
\label{eq_Dirac_time_ind_charge_conj}
\end{equation}%
where $\tilde{\psi}_{c}=\alpha \tilde{\psi}^{\ast }$ and $H_{c}$ is given by
\begin{equation}
H_{c}=c\alpha \,p+\beta mc^{2}-I\,V_{t}-\alpha V_{sp}+\beta V_{s}+i\beta {%
\gamma ^{5}}V_{p}\ .
\end{equation}%
In terms of the potentials $\Delta $ and $\Sigma $, this Hamiltonian reads
\begin{equation}
H_{c}=c\alpha \,p+\beta mc^{2}-\alpha V_{sp}-\frac{I+\beta }{2}\Delta -\frac{%
I-\beta }{2}\Sigma +i\beta \gamma ^{5}V_{p}\ .
\end{equation}%
One sees that the charge-conjugation operation changes the sign of the
energy and of the potentials $V_{t}$, $V_{sp}$ and $V_{p}$. In turn, this
means that $\Sigma $ turns into $-\Delta $ and $\Delta $ into $-\Sigma $.
Therefore, to be invariant under charge conjugation, the Hamiltonian must
contain only a scalar potential.

\subsection{Chiral transformation}

\label{SubSec:chiral}

The chiral operator for a Dirac spinor is the matrix $\gamma^5$, and we will call
``chiral transformation'' the transformation associated with it. Thus,
the transformed spinor is given by $\Psi_\chi=\gamma^5\Psi$ and the transformed
Hamiltonian $H_\chi=\gamma^5H\gamma^5$. Since $%
\gamma^5$ anticommutes with $\beta$, the time-independent chiral-transformed
Dirac equation is
\begin{equation}  \label{eq_Dirac_time_ind_chiral}
H_\chi\tilde\psi_\chi=E\tilde\psi_\chi\ ,
\end{equation}
where $H_\chi$ is given by
\begin{equation}  \label{Hamilt_chi}
H_\chi =c\alpha\,p-\beta mc^2 +I\,V_t+\alpha V_{sp}-\beta V_s+i\beta\gamma^5
V_p
\end{equation}
or
\begin{equation}
H_\chi = c\alpha\,p-\beta mc^2 +\alpha V_{sp}+\frac{I+\beta}2\Delta+ \frac{%
I-\beta}2\Sigma+i\beta\gamma^5 V_p
\end{equation}
in terms of $\Sigma$ and $\Delta$. This means that the chiral transformation
changes the sign of the mass and of the scalar and pseudoscalar potentials,
thus turning $\Sigma$ into $\Delta$ and vice-versa. A chiral-invariant
Hamiltonian needs to have zero mass and $V_s$ and $V_p$ zero everywhere.

\subsection{Equations of motion}

\label{SubSec:Eq_motion}

The space component of the 2-vector potential in (\ref{def_V}) can be
absorbed into the wave function by defining a new spinor $\psi $ such that
\begin{equation}
\tilde{\psi}=e^{-i\Lambda }\psi \ ,  \label{def_psi}
\end{equation}%
in which $\Lambda =(1/\hbar c)\,\int^{x}V_{sp}(x^{\prime })\,dx^{\prime }$,
since we have $H\tilde{\psi}=e^{-i\Lambda }(H-\alpha V_{sp})\psi $. From this
point on, we will refer to $V_{t}$ as simply a vector potential, following
the common usage of this term (usually denoted by $V_{v}$). If we now write
the spinor $\psi $ in terms of its components,
\begin{equation}
\psi =\left(
\begin{array}{c}
\psi _{+} \\
\psi _{-} \\
\end{array}%
\right) \ ,  \label{def_psi_comp}
\end{equation}%
the Dirac equation gives rise to two coupled first-order
equations for $\psi _{+}$ and $\psi _{-}$:
\begin{eqnarray}
\label{eq_mot_1}
-i\hbar c\psi _{-}^{\prime }+mc^{2}\psi _{+}+\Sigma \psi _{+}-i\,V_{p}\psi
_{-} &=&E\psi _{+}   \\
\label{eq_mot_2}
-i\hbar c\psi _{+}^{\prime }-mc^{2}\psi _{-}+\Delta \psi _{-}+i\,V_{p}\psi
_{+} &=&E\psi _{-}\ ,
\end{eqnarray}%
\noindent where the prime denotes differentiation with respect to $x$. In
terms of $\psi _{+}$ and $\psi _{-}$ the spinor is normalized as $%
\int_{-\infty }^{+\infty }dx\left( |\psi _{+}|^{2}+|\psi _{-}|^{2}\right) =1$%
, so that $\psi _{+}$ and $\psi _{-}$ are square integrable functions. It is
clear from the pair of coupled first-order differential equations (\ref%
{eq_mot_1}) and (\ref{eq_mot_2}) that $\psi _{+}$ and $\psi _{-}$ have
opposite parities if the Dirac equation is covariant under $x\rightarrow -x$.

Under the charge-conjugation and chiral transformations the spinor (\ref%
{def_psi_comp}) becomes
\begin{equation}  \label{psi_comp_ch_conj}
\psi_c=\alpha\psi^*=\left(%
\begin{array}{c}
\psi_{-}^* \\
\psi_{+}^* \\
\end{array}
\right)\
\end{equation}
and
\begin{equation}
\label{psi_comp_chiral}
\psi_\chi=\gamma^5\psi=\left(%
\begin{array}{c}
\psi_{-} \\
\psi_{+} \\
\end{array}
\right)\ ,
\end{equation}
respectively.

\subsection{The Sturm-Liouville problem and isolated solutions}

\label{SubSec:Sturm-Liouville}

Using the expression for $\psi_{-}$ obtained from (\ref{eq_mot_2}) with $%
E\neq -mc^{2}+\Delta $, \textit{viz.}
\begin{equation}
\psi_{-} =-i\frac{\hbar c\psi_{+} ^{\prime }-V_{p}\psi_{+} }{E+mc^{2}-\Delta
}  \label{psi-_gen}
\end{equation}
\noindent and inserting it in (\ref{eq_mot_1}) one arrives at the following
second-order differential equation for $\psi_{+}$:
\begin{equation}
-\hbar ^{2}c^{2}\psi_{+} ^{\prime \prime }+\hbar c\Delta ^{\prime } \frac{%
V_{p}\psi_{+}-\hbar c\psi_{+}^{\prime}}{E+mc^{2}-\Delta }+\left[
V_{p}^{2}+\hbar cV_{p}^{\prime } -( E-mc^{2}-\Sigma)( E+mc^{2}-\Delta) %
\right] \psi_{+} =0\ .
\label{2_order_gen_psi+}
\end{equation}

\noindent In a similar way, using the expression for $\psi_{+} $ obtained
from (\ref{eq_mot_1}) with $E\neq mc^{2}+\Sigma $, \textit{viz.}
\begin{equation}
\psi_{+} =-i\,\frac{\hbar c\psi_{-} ^{\prime }+V_{p}\psi_{-} }{%
E-mc^{2}-\Sigma }  \label{psi+_gen}
\end{equation}
\noindent and inserting it in (\ref{eq_mot_2}) one arrives at the following
second-order differential equation for $\psi_{-} $:
\begin{equation}
-\hbar ^{2}c^{2}\psi_{-} ^{\prime \prime }-\hbar c\Sigma ^{\prime } \frac{%
V_{p}\psi_{-}+\hbar c\psi_{-}^{\prime}}{E-mc^{2}-\Sigma}+\left[
V_{p}^{2}-\hbar cV_{p}^{\prime } -( E-mc^{2}-\Sigma)( E+mc^{2}-\Delta) %
\right] \psi_{-} =0  \label{2_order_gen_psi-}
\end{equation}
\noindent For $\Delta =0$ with $E\neq -mc^{2}$, (\ref{psi-_gen}) and (\ref%
{2_order_gen_psi+}) reduce to
\begin{equation}
\psi_{-} =-i\,\frac{\hbar c\psi_{+} ^{\prime }-V_{p}\psi_{+} }{E+mc^{2}}\ ,
\label{psi-_Delta=0}
\end{equation}
\begin{equation}
-\frac{\hbar ^{2}}{2m}\,\psi_{+} ^{\prime \prime }+\left[ \frac{E+mc^{2}}{%
2mc^{2}}\,\Sigma +\frac{V_{p}^{2}}{2mc^{2}}+\frac{\hbar V_{p}^{\prime }}{2mc}%
\right] \psi_{+} =\frac{E^{2}-m^{2}c^{4}}{2mc^{2}}\psi_{+}\ ,
\label{2_order_Delta=0_psi+}
\end{equation}
\noindent and for $\Sigma =0$ with $E\neq mc^{2}$, (\ref{psi+_gen}) and (%
\ref{2_order_gen_psi-}) reduce to
\begin{equation}
\psi_{+} =-i\,\frac{\hbar c\psi_{-} ^{\prime }+V_{p}\psi_{-} }{E-mc^{2}}\ ,
\label{psi+_Sigma=0}
\end{equation}
\begin{equation}
-\frac{\hbar ^{2}}{2m}\,\psi_{-} ^{\prime \prime }+\left[ \frac{E-mc^{2}}{%
2mc^{2}}\,\Delta +\frac{V_{p}^{2}}{2mc^{2}}-\frac{\hbar V_{p}^{\prime }}{2mc}%
\right] \psi_{-} =\frac{E^{2}-m^{2}c^{4}}{2mc^{2}}\psi_{-}\ .
\label{2_order_Sigma=0_psi-}
\end{equation}
\noindent Either for $\Delta =0$ with $E\neq -mc^{2}$ or $\Sigma =0$ with $%
E\neq mc^{2}$ the solution of the relativistic problem is mapped into a
Sturm-Liouville problem in such a way that solution can be found by
solving a Schr\"{o}dinger-like problem. If one
considers potentials $V_{t}$ and $V_{s}$ quadratic in $x$ with $V_{t}=\pm
V_{s}$ and a potential $V_{p}$ linear in $x$ one obtains Schr\''{o}%
dinger-like equations for the harmonic oscillator potential.

The solutions for $\Delta =0$ with $E=-mc^{2}$ and $\Sigma =0$ with $%
E=mc^{2} $, excluded from the Sturm-Liouville problem, can be obtained
directly from the original first-order equations (\ref{eq_mot_1}) and (\ref%
{eq_mot_2}). They are
\begin{equation}
\left.
\begin{array}{l}
\displaystyle\psi_{+} =\psi_{+} ^{(0)}\exp \left[ +\int^{x}dy\,\frac{V_{p}(y)%
}{\hbar c}\right] \\
\\
\displaystyle\psi_{-} ^{\prime }+\frac{V_{p}}{\hbar c}\,\psi_{-} =-\frac{i}{%
\hbar c}\left( \Sigma +2mc^{2}\right) \psi_{+}%
\end{array}
\right\} ,\mathrm{\;for\;}\Delta =0\mathrm{\;with\;}E=-mc^{2}
\label{isol_sol_Delta=0}
\end{equation}
\noindent and
\begin{equation}
\left.
\begin{array}{l}
\displaystyle\psi_{-} =\psi_{-} ^{(0)}\exp \left[ -\int^{x}dy\,\frac{V_{p}(y)%
}{\hbar c}\right] \\
\\
\displaystyle\psi_{+} ^{\prime }-\frac{V_{p}}{\hbar c}\,\psi_{+} =-\frac{i}{%
\hbar c}\left( \Delta -2mc^{2}\right) \psi_{-}%
\end{array}
\right\} ,\mathrm{\;for\;}\Sigma =0\mathrm{\;with\;}E=mc^{2}
\label{isol_sol_Sigma=0}
\end{equation}
\noindent where $\psi_{+}^{(0)}$ and $\psi_{-}^{(0)}$ are normalization
constants. Since, by parity conservation, $V_p$ must be an odd function of $%
x $, its integral must be an even function of $x$. Thus, if the integral
of $V_p$ has a definite sign, its overall sign
can determine whether the isolated solutions (\ref{isol_sol_Delta=0})
and (\ref{isol_sol_Sigma=0}) are bound states. In particular, if $V_p$ and $%
\Sigma$ or $\Delta$ are powers of $x$, one sees from each pair of equations (%
\ref{isol_sol_Delta=0}) and (\ref{isol_sol_Sigma=0}) that, in general, one
of the spinor components has to be zero, either $\psi_{+}$ in the first case
(sign of the $V_p$ coefficient positive) or $\psi_{-}$ in the last case
(sign of the $V_p$ coefficient negative). This is the case of the harmonic oscillator
potentials, as we will see in the next section. However, it is possible find solutions
with the $V_p$ coefficient negative in the first case ($\Delta =0$, $E=-mc^{2}$) and with
the $V_p$ coefficient positive in the second case ($\Sigma =0$, $E=mc^{2}$),
provided a certain relation between the coefficients of all the potentials
is satisfied, as we will see later.

\subsection{The nonrelativistic limit}

\label{SubSec:nonrel_limit}

In the nonrelativistic limit, $E\sim mc^{2}$, Eq. (\ref{2_order_gen_psi+}) becomes
\begin{equation}
\label{eq_mot_nr2a}
\frac{\hbar^2c^2}{2mc^2-\Delta}\left[ -\psi_{+}^{\prime\prime}-\frac{%
\Delta^{\prime}}{(2mc^2-\Delta)}(\psi_{+}^{\prime}-\frac{V_p}{\hbar c}%
\psi_{+}) +\frac{V_p^{\prime}}{\hbar c}\psi_{+}+\frac{V_p^2}{\hbar^2 c^2}%
\psi_{+}\right]+\Sigma\psi_{+}=\mathcal{E}\psi_{+} \ ,
\end{equation}
where $\mathcal{E}=E-mc^2$. Since, in the nonrelativistic regime, for the
range of values of $%
x $ in which the wave function is not negligible, $|\Delta(x)|\ll mc^2$, from Eq. (%
\ref{psi-_gen}) we see that $\psi_{-} $ is of order $v/c\ll 1$ relative to
$\psi_{+}$, provided, of course, one has also $|V_p(x)|\ll mc^2$.
On the other hand, since in these conditions $2mc^2-\Delta\sim 2mc^2$ and
the $\Delta'$ term in Eq.~(\ref{eq_mot_nr2a}) is suppressed relative to the other
terms, we also see that $\psi_{+}$
obeys a Schr\"{o}dinger equation with binding energy equal to $\mathcal{E}%
=E-mc^2$.

We shall consider now the nonrelativistic limit in the special cases $%
\Delta=0$ and $\Sigma=0$, which we will also consider later in the paper
when we solve the Dirac equation for harmonic oscillator potentials.

When $\Delta=0$, Eq.~(\ref{eq_mot_nr2a}) becomes
\begin{equation}
 \label{eq_mot_nr2a_Delta=0}
-\frac{\hbar^2}{2m}\psi_{+}^{\prime\prime}+\bigg(\frac{\hbar}{2mc}%
V_p^{\prime}+ \frac{V_p^2}{2mc^2}+\Sigma\bigg)\psi_{+}= \mathcal{E}%
\psi_{+} \ .
\end{equation}

\noindent Here we clearly see that $\Sigma$ plays the role of a binding
potential in the nonrelativistic limit and that $V_p$ gives rise to
effective binding potentials proportional to $V_p^{\prime}$ and $V_p^2$.
This means that even a pseudoscalar potential unbounded from below could be
a confining potential.
Note that if $V_p/(mc^2)$ is of the same order that $\Sigma/(mc^2)$ in a $1/(mc^2)$
expansion within the
classically accessible region (positive kinetic energy) where the wave
function is not negligible, and if its variation over this region is such
that $\hbar/(2mc)V_p^{\prime}\sim 0$ ($V_p$ changes very little over a
distance of the order of a Compton wavelength), $V_p$ gets suppressed relative to $%
\Sigma$.

For the case $\Sigma=0$ and
if $|\Delta|/(mc^2)$ is very small in the classically accessible
region, Eq.~(\ref{eq_mot_nr2a}) becomes
\begin{equation}
\label{eq_mot_nr2a_Sigma=0}
-\frac{\hbar^2}{2m}\psi_{+}^{\prime\prime}+\bigg(\frac{\hbar}{2mc}%
V_p^{\prime}+ \frac{V_p^2}{2mc^2}\bigg)\psi_{+}=\mathcal{E}\psi_{+} \ .
\end{equation}
\noindent If $V_p=0$ or if, as explained before, the $V_p^2$ and $V'_p$ terms are
higher-order terms in a $1/(mc^2)$ expansion within the classically accessible region,
Eq. (\ref{eq_mot_nr2a_Sigma=0}) becomes a free particle Schr\"odinger equation.
This is in agreement with what was found in
Ref.~\cite{nosso}, namely that when $\Sigma=V_p=0$ and $\Delta$ is a
three-dimensional harmonic oscillator potential, there is no nonrelativistic
limit. In this case we were able to show that this is true in a more general
framework. A model for a relativistic particle in which the vector and
scalar potentials are such that $\Sigma=0$, and where the pseudoscalar
potential is suppressed, is intrinsically relativistic, as far as bound
states are concerned.

This last result deserves some more comments. The fact that in a Dirac
equation one can have Lorentz scalar and pseudoscalar potentials leads to
remarkable results that are at odds with what is known from nonrelativistic
quantum mechanics, where the Lorentz structure plays no role, since one has
only potentials which couple to the energy, \textit{i.e.}, behave as time
components of a relativistic vector. Among these results is the fact that
the Dirac equation is not invariant under a simultaneous shift of the
energy, the scalar potential, and the pseudoscalar potential. It has already
been verified that a constant added to the pseudoscalar screened Coulomb
potential \cite{asc10} or to the pseudoscalar inversely linear potential \cite%
{asc100} is physically relevant and plays a
crucial role in ensuring the existence of bound states. As an example of
another notable result, it is well known that a confining potential for a
nonrelativistic fermion cannot confine it in the
relativistic regime when the potential is considered as a Lorentz vector
(see, \textit{e.g.}, \cite{tha}), simply because there is pair creation and the
single-particle picture does not hold anymore.
It is surprising that the converse is also true, \textit{%
i.e.}, that relativistic binding potentials may not bind in the
nonrelativistic limit. This is the case of $V_{t}=-V_{s}$ ($\Sigma =0$)
referred before, and is basically due to the fact that vector potentials
couple in a different way than scalar potentials in the Dirac equation,
whereas there is no such distinction in the Schr\"{o}dinger equation.
Actually, as we have seen before, $\Sigma$, which is the sum of a vector and a
scalar potential, plays the role of a binding potential in the
nonrelativistic limit, whereas a weak enough $\Delta$
cannot bind a fermion with nonrelativistic energy if $\Sigma=0$.
Note that for potentials going to zero at infinity, like mean-field nuclear potentials,
there are no bound states when $\Sigma=0$, even if $\Delta$ is not small compared with
fermion rest energy \cite{gin3}-\cite{alb2}.

Regarding antiparticles, an attractive vector potential for a particle is,
of course, repulsive for its corresponding antiparticle, and vice versa.
However, the attractive (repulsive) scalar potential for particles is also
attractive (repulsive) for antiparticles. This is expressed in the change $%
\Delta\to -\Sigma$ and $\Sigma\to -\Delta$ under charge conjugation, as
referred in Sec.~\ref{SubSec:charge_conj}. This means that an antiparticle
with nonrelativistic energies ($E\approx -mc^2$) will not be bound when $\Delta=0$,
provided $V_p$ is small enough.

\section{The relativistic harmonic oscillator}

\label{Sec:rel_h_oscillator}

Let us consider

\begin{equation}
\Sigma =\frac{1}{2}k_{1}x^{2},\qquad \Delta =0,\qquad V_{p}=k_{2}x
\label{Sigma_OH}
\end{equation}

As we have seen in the last section, the chiral transformation performs the
changes $\Delta\to\Sigma$, $\Sigma\to\Delta$, $m\to -m$ and $V_p\to -V_p$.
Morever, since $\gamma^5$ interchanges the upper and lower components (see (%
\ref{psi_comp_chiral})), the resulting pair of transformed equations of
motion are formally the same, so that their solutions have the same energy
eigenvalues. This symmetry can be clearly seen from the two equation pairs
(\ref{psi-_gen})-(\ref{2_order_gen_psi+}) and (\ref{psi+_gen})-(\ref{2_order_gen_psi-})
as well as from the isolated solutions
(\ref{isol_sol_Delta=0}) and (\ref{isol_sol_Sigma=0}) which are converted
into each other by this kind of transformation.
This means that one can restrict the discussion to the $\Sigma$
case ($\Delta =0$). The results for the case when $\Delta=\frac{1}{2}%
k_{1}x^{2},\ \Sigma =0,\ V_{p}=k_{2}x$ can be obtained immediately by just
changing the sign of $m$ and of $k_2$ in the relevant expressions.

The Dirac spinor corresponding to the isolated solution with $E=-mc^{2}$ is
obtained from (\ref{isol_sol_Delta=0}). Only for $k_{2}>0$ there is a
normalizable Dirac spinor and, as commented before, the upper component
vanishes whereas the lower component is given by $\psi_{-}
=\psi_{-}^{(0)}\exp \left(- k_{2}x^{2}/(2\hbar c)\right) $, regardless the
value of $k_2$. For $E\neq -mc^{2}$, Eq. (\ref{2_order_Delta=0_psi+})
takes the form
\begin{equation}
-\frac{\hbar ^{2}}{2m}\psi_{+} ^{\prime \prime }+\frac{1}{2}K x^{2}\psi_{+} =%
\tilde{E}\psi_{+}  \label{2_order_Delta=0_OH_psi+}
\end{equation}
\noindent where
\begin{eqnarray}
\label{def_K}
K&=&\frac{1}{m c^{2}}\left( \frac{E+mc^{2}}{2}\,k_{1}+k_{2}^{2}\right) \\
\label{def_EE}
\tilde{E}&=&\frac{E^{2}-m^{2}c^{4}}{2mc^{2}}-\frac{\hbar k_{2}}{2mc} \ .
\end{eqnarray}

\noindent The well-behaved solution for (\ref{2_order_Delta=0_OH_psi+}),
with $K$ necessarily real and positive, is the well-known solution of the
Schr\"{o}dinger equation for the nonrelativistic harmonic oscillator (see,
\textit{e.g.}, \cite{sch}):
\begin{eqnarray}
\psi_{+} &=&N_{n}H_{n}\left( \lambda x\right) e^{-\lambda ^{2}x^{2}/2}
\label{psi+_Delta=0_OH} \\
&&  \nonumber \\
\tilde{E} &=&\left( n+\frac{1}{2}\right) \hbar \sqrt{\frac{K}{m}}
\label{Delta=0_OH_EE}
\end{eqnarray}
\noindent where $n=0,1,2,\ldots $, $N_{n}$ is a normalization constant, $%
H_{n}\left( \lambda x\right) $ is a $n$-th degree Hermite polynomial and
\begin{equation}
\lambda =\left( \frac{mK}{\hbar ^{2}}\right) ^{1/4}  \label{def_lambda}
\end{equation}

The lower component of the Dirac spinor is obtained from (\ref%
{psi-_Delta=0}). When one uses the recursion relations for the Hermite
polynomials (see, e.g., \cite{abra}) one gets
\begin{equation}
\psi_{-} =\frac{i\hbar c\lambda N_{n}}{E+mc^{2}}\left[ H_{n+1}\left( \lambda
x\right) -\left( 1-\frac{k_{2}}{\hbar c\lambda ^{2}}\right) \left( \lambda
x\right) H_{n}\left( \lambda x\right) \right] e^{-\lambda ^{2}x^{2}/2}
\label{psi-_Delta=0_OH}
\end{equation}
\noindent Since the Hermite polynomial $H_{n}\left( \lambda x\right) $ has $%
n $ distinct zeros one may conclude that $\psi_{+} $ has $n$ nodes, and the
expression for $\psi_{-} $ suggests that it has $n\pm 1$ nodes, depending on
the sign $k_2$. In fact, if $k_1=0$, this last expression reduces to
\begin{equation}
\psi_{-} =\frac{i\hbar c\lambda N_{n}}{E+mc^{2}}\left[ H_{n+1}\left( \lambda
x\right) -\left( 1-\frac{k_{2}}{|k_2|}\right) \left( \lambda x\right)
H_{n}\left( \lambda x\right) \right] e^{-\lambda ^{2}x^{2}/2} \ ,
\label{psi-_Delta=0_k1=0}
\end{equation}
showing that $\psi_{-}$ is proportional to $H_{n+1}( \lambda x)$ if $k_2$ is
positive. If $k_2<0$, the recursion relations of Hermite polynomials imply
that $\psi_{-}$ is proportional to $H_{n-1}( \lambda x)$ (see \cite{abra}).

From Eqs. (\ref{def_K}), (\ref{def_EE}) and (\ref{Delta=0_OH_EE}) we can get
the quantization condition for the Dirac eigenenergies:
\begin{equation}
E^{2}-m^{2}c^{4} =\left( 2n+1\right)\hbar c\, \sqrt{\frac{E+mc^{2}}{2}%
k_{1}+k_{2}^{2}}+\hbar c\,k_{2}  \label{eigen_eq_Delta=0}
\end{equation}
\noindent The nonrelativistic limit is reached when $\hbar^2 |k_1|/(m^3
c^4)\ll 1$ and $\hbar |k_2|/(m^2 c^3)\ll 1$. In this case $E\sim mc^2$ for
small quantum numbers, and Eq. (\ref{eigen_eq_Delta=0}) becomes
\begin{equation}
E-mc^{2}=\left( n+\frac{1}{2}\right) \hbar \,\sqrt{\frac{k_{1}}{m}+\left(
\frac{k_{2}}{mc}\right) ^{2}}+\frac{\hbar k_2}{2mc}  \label{non_rel_eigen_eq}
\end{equation}
\noindent so that $k_{1}$ is restricted to $k_{1}>-k_{2}^{2}/(mc^{2})$.

In general there is no requirements on the signs of $k_{1}$ and $k_{2}$, except that,
as stated before, $K>0$, and therefore, from (\ref{def_K}),
the Dirac eigenenergies corresponding to the bound-state
solutions must be within the limits
\begin{eqnarray}
E &>&-mc^{2}-2\,\frac{k_{2}^{2}}{k_{1}},\qquad \mathrm{for\quad }k_{1}>0
\label{limit_values_k1_p} \\
&&  \nonumber \\
E &<&-mc^{2}+2\,\frac{k_{2}^{2}}{|k_{1}|},\qquad \mathrm{for\quad }k_{1}<0
\label{limit_values_k1_n}
\end{eqnarray}
\noindent For $k_{1}=0$, when there is a pure pseudoscalar potential linear
in $x$ (the two-dimensional Dirac oscillator), Eq.~(\ref{eigen_eq_Delta=0})
reduces to
\begin{equation}
E^{2}=m^{2}c^{4}+\hbar c\left[ \left( 2n+1\right) |k_{2}|+k_{2}\right]\ .
\label{energy_eigen_k1=0}
\end{equation}
\noindent Note that $n\geq 0$ for $k_2>0$ and $n\geq 1$ for $k_2<0$,
because for the latter case the lower component is
proportional to a Hermite polynomial of degree $n-1$. This result, together
with Eq. (\ref{energy_eigen_k1=0}), allows us to conclude that the Dirac
eigenvalues are given by $E=\pm E_{0}$, where
$E_{0}=\sqrt{m^{2}c^{4}+2(n+1)\hbar c|k_{2}|}>mc^{2}$, $n=0,1,\ldots$,
irrespective of the sign of $k_{2}$. This means that \textit{the spectrum is independent
of the sign of $k_2$}. The solutions $E=mc^{2}$ and $E=-mc^{2}$ correspond
to isolated solutions for $k_2<0$ and $k_2>0$, respectively (see Eqs. (\ref%
{isol_sol_Delta=0}) and (\ref{isol_sol_Sigma=0})). However, the spinors do
depend on the sign of $k_2$, because different signs induce different node
structures, as can be seen from (\ref{psi-_Delta=0_k1=0}). Since the
charge-conjugation transformation changes the sign of $k_2$, it connects both kinds
of solutions, but each can have both positive and negative energies. Thus a
complete set of states for a certain $k_2$ includes both particle (positive
energy) and antiparticle (negative energy) states but they are not related
to each other by a charge-conjugation transformation. The spectral gap between these
solutions is $2\sqrt{m^2c^{4}+2\hbar c|k_{2}|}$ and does not disappear when $%
m=0$. Therefore, there is no room for transitions from positive- to
negative-energy solutions and Klein's paradox never comes into play. There are no
scattering states when $k_{1}=0$, but for $k_{1}\neq 0$ the continuum states
are those ones which have energies beyond the limits expressed by (\ref%
{limit_values_k1_p})-(\ref{limit_values_k1_n}), for which $K\leq 0$. As far as
the chiral transformation is concerned, one can note that $\gamma^5$ acts
much in the same way as the charge-conjugation transformation, except that
it does not change the sign of the energy and changes the mass sign. Therefore,
the chiral transformation connects solutions with the same
energy sign but that have opposite $k_2$ signs.

When $k_{2}=0$, \textit{i.e.}, no pseudoscalar potential, there is no
isolated bounded solutions and the quantization condition expressed by Eq. (%
\ref{eigen_eq_Delta=0}) takes the form
\begin{equation}
\left( E-mc^{2}\right) \sqrt{\left( E+mc^{2}\right) \mathrm{sgn}(k_{1})}=%
\mathrm{sgn}(k_{1})\left( n+\frac{1}{2}\right) \hbar c\sqrt{2|k_{1}|}
\label{eigen_eq_Delta=0_k2=0}
\end{equation}
\noindent where $\mathrm{sgn}(k_{1})$ stands for the sign function. This
means that when $k_{1}>0$ there are only bound states for fermions with $%
E>mc^{2}$ and those bounded solutions are kept apart from the
negative-energy continuum ($E<-mc^{2}$) by an energy gap greater than $%
2mc^{2}$. On the other hand, for $k_{1}<0$ there are only bound states for
antifermions with $E<-mc^{2}$ and those bounded solutions also do not mix
with continuum states with $E>-mc^{2}$. Therefore, the positive-energy bound
states of fermions never sink into the Dirac sea of negative energies, and
the negative-energy bound states of antifermions never reach the upper
continuum, so again there is no danger of reaching the conditions for
Klein's paradox. It is important to note that the solutions with $k_1<0$
\textit{are not} the antiparticles of the solutions with $k_1>0$ because
charge conjugation changes $\Sigma$ into $-\Delta$ and vice-versa. This
means that the antiparticles of the states with $\Sigma=\frac{1}{2}%
k_{1}x^{2},\ \Delta =0,\ V_{p}=0$ and $k_1>0$ are the negative-energy
solutions of the relativistic system with $\Delta=-\frac{1}{2}k_{1}x^{2},\
\Sigma =0,\ V_{p}=0$, with $k_1$ positive.

To conclude, let us present, as an example of the rule stated at the
beginning of the section, the eigenvalue equation for the relativistic
harmonic oscillator when $\Delta =\frac{1}{2}k_{1}x^{2},\ \Sigma =0,\
V_{p}=k_{2}x$:
\begin{equation}
E^{2}-m^{2}c^{4}=\left( 2\tilde{n}+1\right)\hbar c\, \sqrt{\frac{%
E-mc^{2}}{2}k_{1}+k_{2}^{2}}-\hbar c\, k_{2}\ .
\label{eigen_eq_Sigma=0}
\end{equation}%
Here $\tilde{n}$ is the number of nodes (the degree of the Hermite
polynomial) for the lower component, since the chiral transformation
interchanges the upper and lower components, so that the second-order
equation to solve in this case is the one for the lower component,
Eq.~(\ref{2_order_Sigma=0_psi-}). The upper
component would be obtained from the corresponding first-order equation
Eq.~(\ref{psi+_Sigma=0}). As a
final remark, one may note that for massless particles, the $\Delta =0$ and $%
\Sigma =0$ oscillators have the same spectrum with the sign of $k_{2}$
reversed.

In 3+1 dimensions the corresponding eigenvalue equations to Eqs.~(\ref%
{eigen_eq_Delta=0}) and (\ref{eigen_eq_Sigma=0}) have a very similar
structure. From \cite{nosso} we see that, for $\Delta=0$, one just adds
the Dirac equation quantum number $\kappa$ to the coefficient of $k_2$
and modifies the coefficient of the square root to include the orbital angular momentum
$l$ of the upper component
\begin{equation}
E^{2}-m^{2}c^4=\left( 4n+2l+3\right)\hbar c\, \sqrt{\frac{E+mc^2}{2}k_1+k_2^2}
+(2\kappa -1)\hbar c\,k_2\ ,
\label{eigen_eq_Delta=0_3+1}
\end{equation}
while for $\Sigma=0$ one uses the orbital angular momentum $\tilde l$ of the lower
component
\begin{equation}
E^{2}-m^{2}c^4=\left( 4\tilde n+2\tilde l+3\right)\hbar c\, \sqrt{\frac{E-mc^2}{2}k_1+k_2^2}
-(2\tilde\kappa -1)\hbar c\,k_2\ ,
\label{eigen_eq_Sigma=0_3+1}
\end{equation}
where $\tilde\kappa=-\kappa$ (see \cite{nosso}). If we compare these equations with
(\ref{eigen_eq_Delta=0}) and (\ref{eigen_eq_Sigma=0}), we see that the main features
are identical, and thus all the main results obtained in the next section remain valid for the
relativistic harmonic oscillator in three spatial dimensions.

\section{Graphical determination of the energy levels}

\label{Sec:Graph_det_e_levels}

We turn now to the solutions of Eq. (\ref{eigen_eq_Delta=0}). In order to
obtain a deeper insight on those solutions for arbitrary $k_{1}$ and $k_{2}$
one has to seek a convenient method for solving this transcendental
equation. If we square Eq. (\ref{eigen_eq_Delta=0}) the resulting equation
is in general a quartic algebraic equation in $E$, which in principle can be
solved analytically. The price paid is that some spurious roots can appear
in this process, although, of course, these can be eliminated by checking
whether they satisfy the original equation. A more instructive procedure is
follow a graphical method, by which one seeks the intersection points of the
functions of energy in Eq. (\ref{eigen_eq_Delta=0}): a parabola on the
left-hand side
\begin{equation}  \label{def_f_p}
f_p(E)= E^{2}-m^{2}c^{4}\ ,
\end{equation}
and a square-root function on the right-hand side
\begin{equation}  \label{def_f_s}
f_s(E)=(2n+1)\hbar c\sqrt{(E+mc^{2})k_{1}/{2}+k_{2}^{2}}+\hbar c k_{2} \ .
\end{equation}
In this way, it is easy to see that there are at most two acceptable
solutions of (\ref{eigen_eq_Delta=0}), since the parabola and the square-root function
can have at most two intersection points. The
construction of the graph of $f_{s}(E)$ is split into three distinct classes
depending on the $k_{1}$ values. Below we discuss these three classes of problems. It
should not be forgotten, though, that the solutions with $E=-mc^{2}$ are not
included in this discussion, since they correspond to isolated solutions for
$k_{2}>0$.

This kind of analysis and the conclusions that will be drawn from it in the following
subsections can be applied
with appropriate modifications to the 3+1 case, and therefore to nucleon and antinucleon
single-particle spectra. This is especially true as far the relations between
the spectra of systems with $\Delta=0$ and $\Sigma=0$ are concerned, since
these are obtained from the general properties of charge-conjugation
and $\gamma^5$ transformations which hold in 3+1 dimensions.

\subsection{The case $k_{1}=0$}

Although we have already solved this case, it is presented here for the sake
of completeness. For $k_{1}=0$, $f_{s}(E)$ is just a nonnegative constant
function as can be seen from Eq. (\ref{energy_eigen_k1=0}), having two
symmetric intersections with the parabola.

\begin{figure}[!ht]
\begin{center}
\includegraphics[width=10cm]{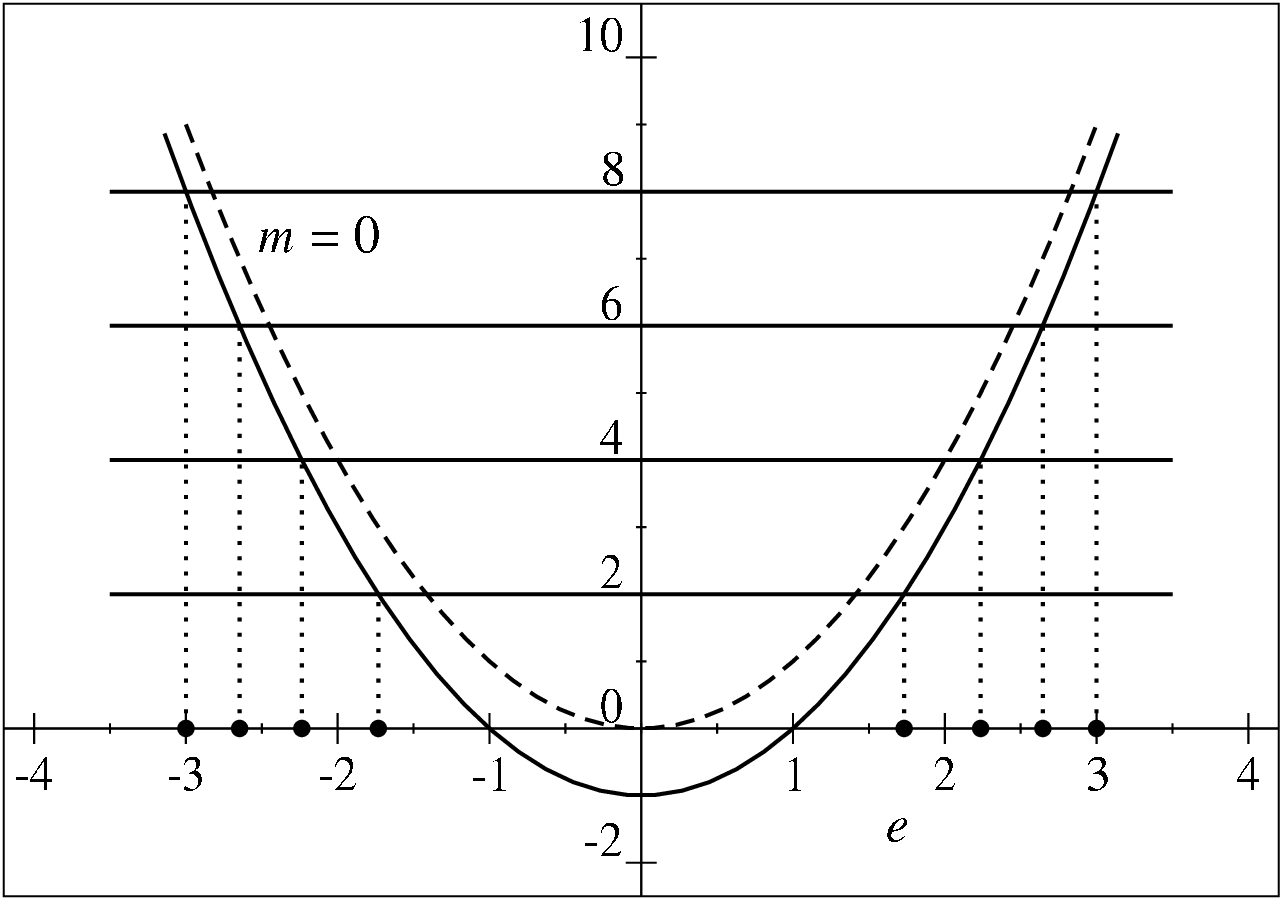}
\end{center}
\caption{Scaled energies $e=E/(mc^2)$ for the first four levels when
$|\protect\kappa_2|=1$ and $%
\protect\kappa_1=0$. The black dots represent the values of energies coming
from the intersection of the parabola and the horizontal lines correspond to
the scaled function $f_{s}(E)/(mc^2)^2$ for $n=0,1,2,3$. The dashed parabola
is the function $f_p(E)$ for $m=0$. In this case, the horizontal lines are
just $f_{s}(E)$ for $n=0,1,2,3$, $|\protect\kappa_2|=1$ in appropriate units
and the horizontal scale is in the units of the energy $E$.\hfill\ }
\label{Fig1}
\end{figure}

\begin{figure}[!ht]
\begin{center}
\includegraphics[width=10cm]{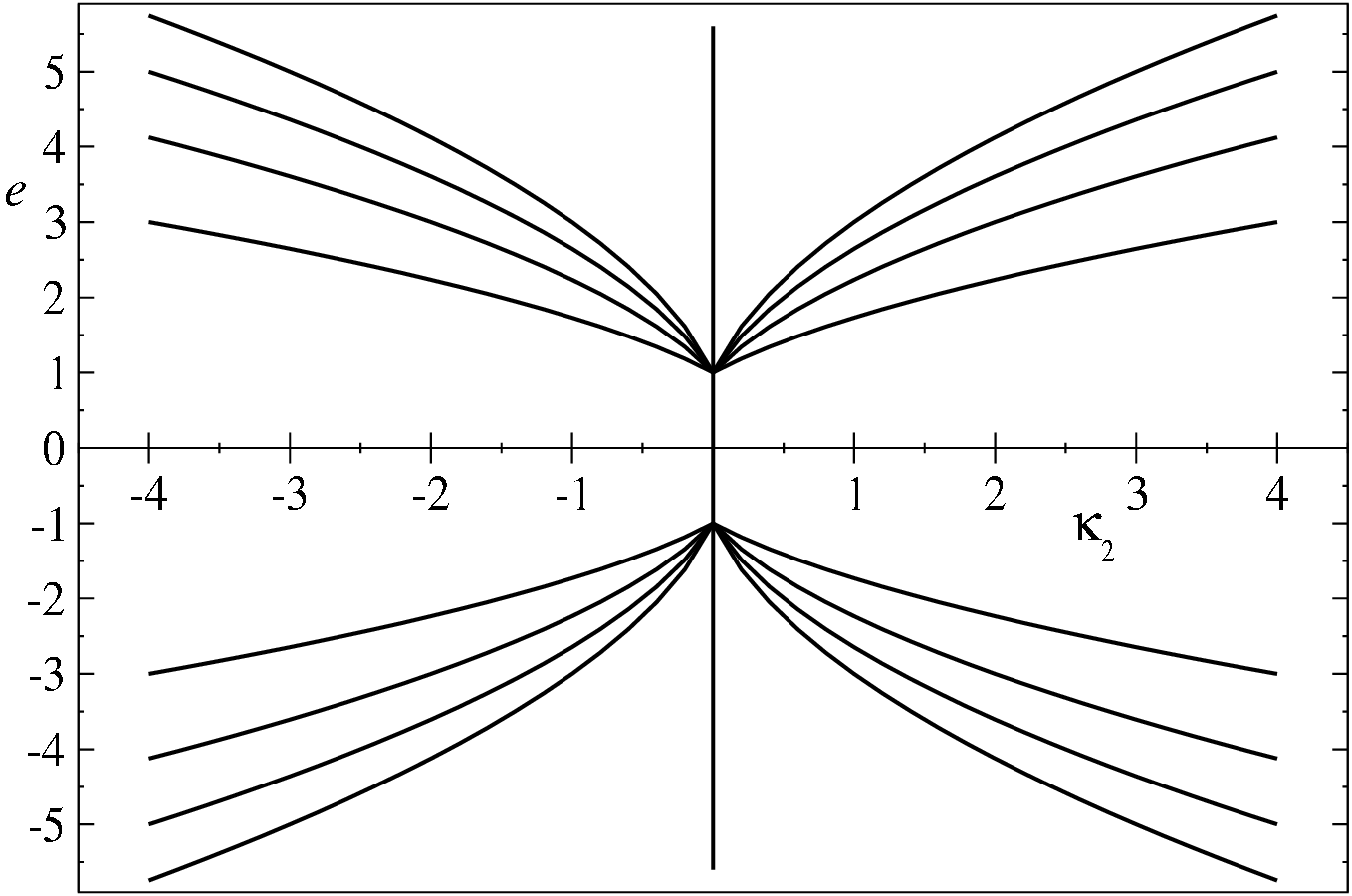}
\end{center}
\caption{Scaled energies $e=E/(mc^2)$ for the first four levels when $\protect\kappa_1=0$ as a
function of $\protect\kappa_2$. For $m=0$ the plots are qualitatively the
same, except that the lines for the different levels begin at the origin
and the vertical scale would be in units of the energy $E$.}
\label{Fig2}
\end{figure}

Fig. \ref{Fig1} and Fig. \ref{Fig2} present the results for the first four
quantum numbers. In these and the following plots we use the scaled
quantities $e=E/(mc^2)$, $\kappa_1=\hbar^2 k_1/(m^3c^4)$ and $\kappa_2=\hbar
k_2/(m^2c^3)$. The first shows how the values of the energy are obtained
from the intersection of the two curves and the second how the eigenvalues
vary with $k_2$. The energies for the zero mass case can also be computed
from the intersections with the dashed parabola in Fig. \ref{Fig1}. From
these figures it is clear the symmetry of the energy levels. Note again that
the isolated solution is not plotted.

\subsection{The case $k_{1}>0$}

When $k_{1}>0$, $f_{s}(E)$ is a monotonically increasing function of $E$.
For each $n$, it can intersect the parabola at one or two points, depending
on the particular values of $k_1$ and $k_2$. To determine the conditions for
this to occur, we have first to determine the minimum of $f_{s}(E)$, which
occurs when the expression under the square root in (\ref{def_f_s}) is zero:
\begin{equation}
E_{\min }=-mc^{2}-2\,\frac{k_{2}^{2}}{k_{1}}\ .  \label{E_min}
\end{equation}
This value sets a lower bound for the energies of all the solutions of Eq. (%
\ref{eigen_eq_Delta=0}). For $E=E_{\min}$, the function $f_{s}(E)$ takes the
value
\begin{equation}
f_{s_{\min }}=\hbar c k_{2} \ ,  \label{f_s_min}
\end{equation}
and $f_{p}$ takes the value
\begin{equation}
f_{p_{\min }}=4\frac{k_{2}^{2}}{k_{1}}\left( \frac{k_{2}^{2}}{k_{1}}%
+mc^{2}\right)  \label{f_p_min}
\end{equation}
When $k_1>0$, the minimum of $f_s(E)$
occurs for negative values of the energy, that is, on the left (and
decreasing) side of the parabola. If $f_{p_{\min }}<f_{s_{\min }} $ there is
only one solution of Eq. (\ref{eigen_eq_Delta=0}) and it has positive
energy. From (\ref{f_s_min}), this can happen only for $k_2>0$, since $%
f_{p_{\min }}>0$. If $f_{s_{\min }}\le f_{p_{\min }}$ there are \textit{%
always} two solutions, one negative and one positive, because $%
f_s(0)=(2n+1)\hbar c\sqrt{mc^{2}k_{1}/{2}+k_{2}^{2}}+\hbar c\,k_{2}$ is always
positive, no matter how large is $|k_2|$, for a positive $k_1$. Furthermore,
the positive and negative solutions are such that $|E|>mc^2$, since, when $%
|E|< mc^2$, $f_s(E)>f_p(E)$ and thus there are no solutions. Note, however,
that for $n=0$ and $k_2$ negative there is a solution with $E=-mc^2$, but
this is an isolated solution out of the Sturm-Liouville problem we are
considering.

The considerations of the previous paragraph can be further elaborated by
analyzing the behavior of the function $F(k_2)$
\begin{equation}  \label{def_F(k2)}
F(k_2)=f_{p_{\min }}-f_{s_{\min }}=\frac{4k_2}{k_1^2} \bigg(%
k_{2}^{3}+mc^{2}k_{1}k_{2}-\frac{\hbar ck_{1}^{2}}{4}\bigg)\ .
\end{equation}
According to what was stated in the previous paragraph, if $F(k_2)>0$ one
has two solutions for each $n$, and if $F(k_2)<0$ there is just one solution
for each $n$. To determine the signs of $F(k_2)$ for each value of $k_2$, we
will need to know the values of $k_2$ which are roots of $F(k_2)=0$, or of
the equation
\begin{equation}
k_{2}^{3}+mc^{2}k_{1}k_{2}-\frac{\hbar ck_{1}^{2}}{4}=0 \ ,
\label{fs_min=fp_min}
\end{equation}
excluding the solution $k_2=0$. Applying Descartes'{} rule of signs to this
last equation, we see that it can only have a positive real solution. The
rule states that the number of positive real roots $n_r$ of an algebraic equation
with real coefficients
\begin{equation}\label{Descartes_rule}
a_{k}z^{k}+\cdots +a_{1}z+a_{0}=0
\end{equation}
is never greater
than the number of changes of signs $n_c$ in the sequence $a_{k},\ldots
,a_{1},a_{0}$ (not counting the null coefficients) and, if less, then differs
from $n_c$ by an even number. This means that $n_r$ can take the values
$n_r=n_c,\, n_c-2,
n_c-4,\,\ldots,n_c-2n\ge 0$, $n=0,1,\ldots$ . In this case, for positive $k_1$,
there is only one such
sign change and therefore there is only a positive and unique solution,
which we shall call $k_{2}^{*}$. The function $F(k_2)$ changes sign at $%
k_2=0 $ and $k_2=k_2^*$, being negative in the interval $0<k_2<k_2^*$ and
positive outside this interval. Therefore, according to what we have said
before, there is only one positive-energy solution when $0<k_2<k_{2}^{*}$
and two solutions (one positive and one negative) when $k_2<0$ or $k_2\ge
k_{2}^{*}$.
If $k_2=0$, the two functions intersect at $E_{\min }=-mc^{2}$, but again,
since this solution was excluded from the Sturm-Liouville problem leading to
the eigenvalue equation (\ref{eigen_eq_Delta=0}), we consider only the
harmonic oscillator solutions with energy greater than $mc^2$.

\begin{figure}[!ht]
\begin{center}
\includegraphics[width=10cm]{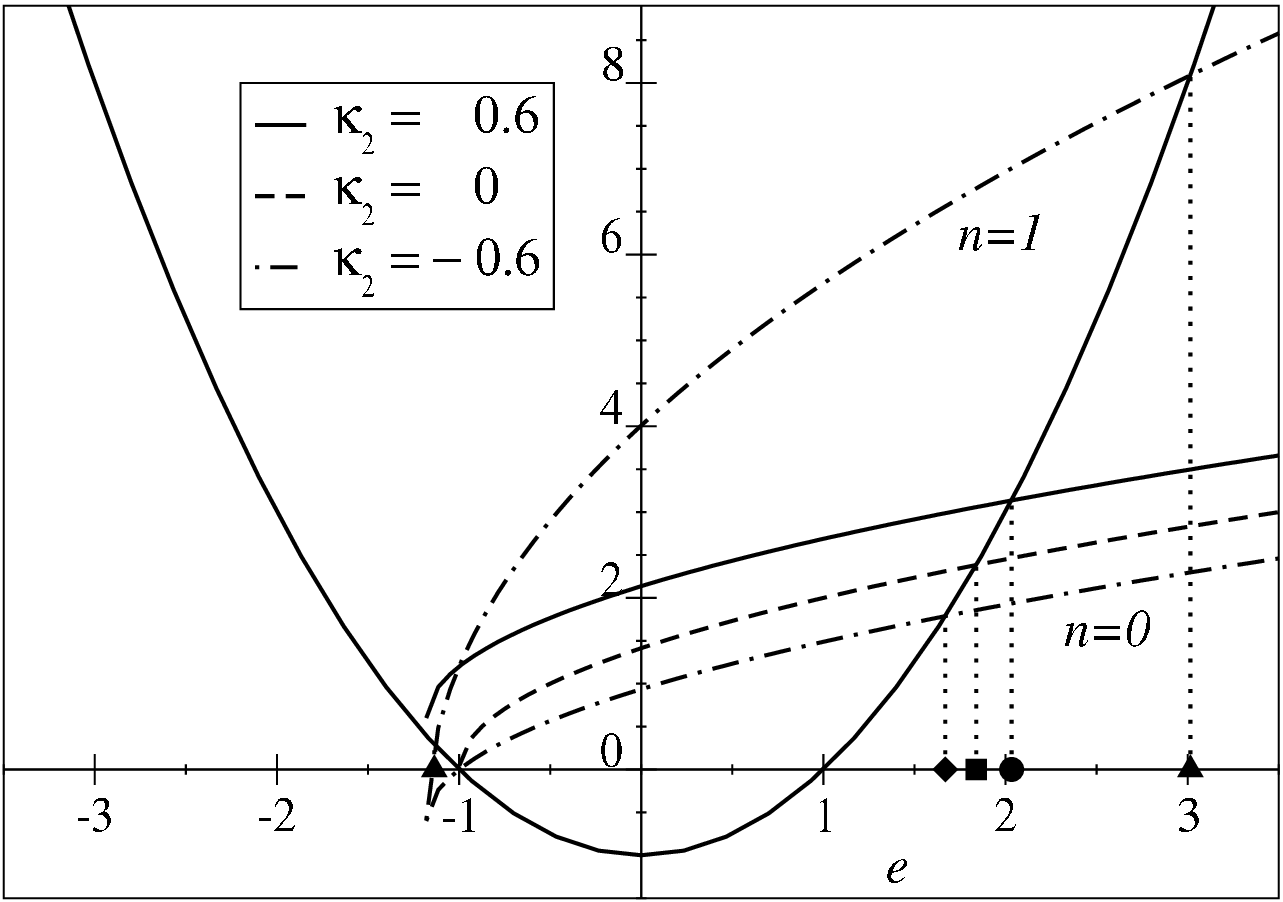}
\end{center}
\caption{Graphical solution of equation (\protect\ref{eigen_eq_Delta=0}) for
$\protect\kappa_1=4$ and three values of $\protect\kappa_2$. The circle,
square and diamond symbols mark the energy solutions for $n=0$ and $\protect%
\kappa_2=0.6, 0,-0.6$, respectively. In the case of $\protect\kappa%
_2=-0.6$, the square root function for $n=1$ is also represented and
the respective energy solutions are marked by the triangle symbols.}
\label{Fig3}
\end{figure}

The graphical representations of $f_{p}$ and $f_{s}$ are shown in Fig.~\ref%
{Fig3} for $m\neq 0$. We plot the square-root function for three values of $%
\kappa_2$ (the scaled $k_2$), for a positive $k_1$. We plot $f_s(E)$ with $%
n=0$ for $\kappa_2=0,\,0.6$ and with $n=0$ and $n=1$ for $\kappa_2=-0.6$.
The value of $\kappa_1$ was chosen such that $F(k_2)$ has a positive value
for $\kappa_2=0.6$. There is always an infinite and unbounded set of
positive-energy states with $E>mc^2$ for any value of $k_2$, and there are
always negative states when $k_2$ is negative.
Note that increasing the value of $n$ does not change qualitatively these
features, as can be seen from the plots of $f_s(E)$ when $k_2$ is negative.

\begin{figure}[!ht]
\begin{center}
\includegraphics[width=10cm]{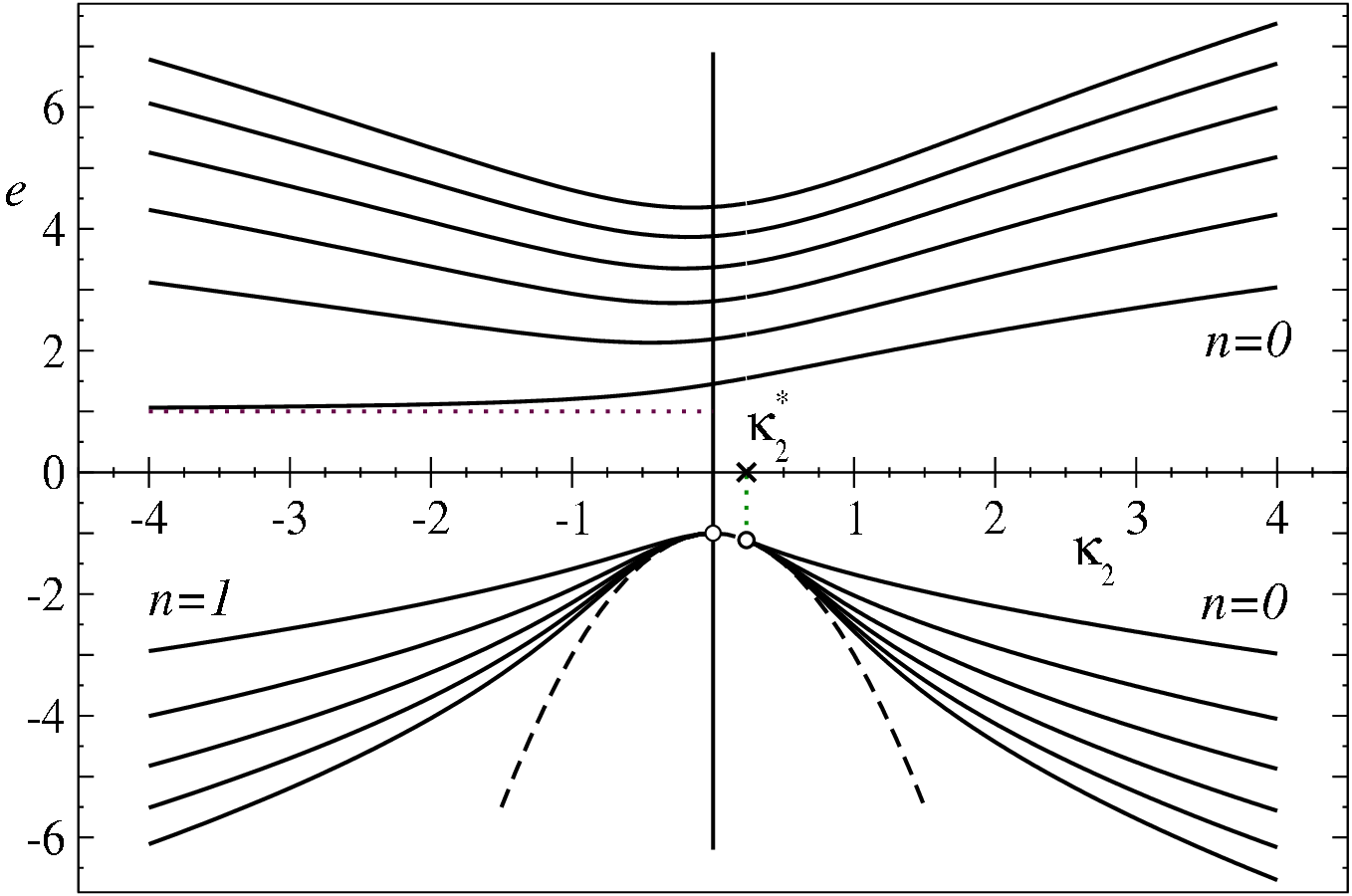}
\end{center}
\caption{Scaled energies for the first six levels when $\protect\kappa_1=1$ as a
function of $\protect\kappa_2$. The $n=0$ level for negative $k_2$ and
negative energy was not plotted because it corresponds to the isolated
solution with $e=-1$ ($E=-mc^2$). The value of $\protect\kappa_2^*=\hbar
k_2^*/(m^2c^3)$ is marked with a cross in the energy axis. The hollow circles
mark the points ($k_2=0$, $e=-1$) and ($k_2=k_2^*$, $e=e_{\rm min}=E_{\rm min}/mc^2$
for which there are no bounded solutions. The
dashed parabola represents the function $-1-2\protect\kappa_2^2/\protect%
\kappa_1$, whose meaning is given in the text.}
\label{Fig4}
\end{figure}

In Fig.~\ref{Fig4} it is clearly seen that for values of $k_2$ between zero
and $k_2^*$ there are only positive-energy solutions. Moreover, one can see
from Eq. (\ref{fs_min=fp_min}) that the value of $k_2^*$ increases with $k_1$%
, meaning that when $k_1\gg |k_2|$ one reaches the limit of the harmonic
oscillator ($k_1>0,\ k_2=0$) where there are only positive bound-state
solutions with $E> mc^2$. Thus, the existence of the pseudoscalar potential
gives rise to negative-energy solutions coming from the Dirac sea. For $%
k_{2}<0$ these kind of solutions always exist. We excluded the isolated
level with $n=0$ for negative $k_2$. Note also that for decreasing negative $%
k_2$ the positive-energy solution for $n=0$ goes to $E=mc^2$, which was the
case we discarded in the discussion of the Dirac oscillator in the previous
section. When $k_{2}=k_{2}^{*}$ there is an infinite set of
degenerate solutions with $E=E_{\min}$, because in this case the intersection
point of $f_s(E)$ with the
negative-energy branch of $f_p(E)$ is the initial (minimum) point of the
square-root function, which of course does not change with $n$ (see Fig.~\ref%
{Fig3}). However, when $E=E_{\min}$, the value of $K$ defined in (\ref{def_K})
is zero, and thus, from (\ref{def_lambda}), $\lambda=0$ and therefore these
solutions do not correspond to bound states (see Eqs. (\ref{psi+_Delta=0_OH})
and (\ref{psi-_Delta=0_OH})). For $k_{2}\gtrsim k_{2}^{*}$
the degeneracy is removed and we have
a very high density of very delocalized states (since $\lambda\gtrsim 0$)
which then decreases with increasing $%
k_2 $. In Fig.~\ref{Fig4} we also plotted the parabola $-mc^2-2k_2^2/k_1$
(in scaled units $-1-2\kappa^2_2/\kappa_1$) which corresponds to the
asymptotic values of the negative-energy states when $n\to\infty$ and thus
represents the set of lower energy bounds of the negative-energy solutions
for each pair of $k_1$ and $k_2$ values. This parabola can be also seen
as delimiting the region in the energy-$k_2$ plane in which $K\leq 0$,
\textit{i.e.}, where there are no bound-state solutions. This is related to
the discussion in the previous section leading to Eqs.~(\ref{limit_values_k1_p}) and
(\ref{limit_values_k1_n}).

As referred before, for a given energy $E<-mc^2$ and positive $k_1$ there
are not bound states for values of $k_2$ in the interval $-\sqrt{%
-(E+mc^2)k_1/2}<k_2<\sqrt{-(E+mc^2)k_1/2}$. In fact, from what was said above
in this section, one can be more precise and state that there are no
negative bound states for $0<k_2< k_2^*$. Since $k_2^*$ is zero when $k_1=0$
and increases with increasing $k_1$, this means that when $k_1=0$ (the Dirac
oscillator) there are no continuum states and when $k_1\to\infty$ (or when $%
k_2\to 0$), all the negative-energy states are continuum states.

\subsection{The case $k_{1}<0$}

For $k_{1}<0$ the function $f_{s}(E)$ decreases monotonically with the
energy, reaching again its minimum when the expression under the square root
in (\ref{def_f_s}) is zero, setting now a maximum value for the energy of
the solutions of Eq.~(\ref{eigen_eq_Delta=0}). The expressions for $E_{\max }
$, $f_{s_{\min }}=f_{s}(E_{\max })$ and $f_{p_{\min }}=f_{p}(E_{\max })$ are
the same as corresponding ones in the previous subsection, respectively Eqs.
(\ref{E_min}), (\ref{f_s_min}) and (\ref{f_p_min}), except that, of course,
now $k_{1}$ is negative.

Noting that $f_{s}$ at $E=-mc^{2}$ has a nonnegative value given by $%
f_{s}(-mc^{2})=(2n+1)\hbar c|k_{2}|+k_{2}$, we see that there is always an
infinite and unbounded set of negative-energy solutions (one for each $n$)
with $E<-mc^{2}$. For $n=0$ and $k_{2}\leq 0$, there is a solution with $%
E=-mc^{2}$ which was already discussed in the previous section. However,
when $k_{1}<0$, one can solve Eq.~(\ref{isol_sol_Delta=0}) when $E=-mc^{2}$
and obtain a bound-state solution with both upper and lower components
different from zero, if and only if $\kappa _{2}=\kappa _{1}/8$. One can
show that this corresponds to a $n=0$ solution given by the functions (\ref%
{psi+_Delta=0_OH}) and (\ref{psi-_Delta=0_OH}) taking the limit $%
E\rightarrow -mc^{2}$.

\begin{figure}[!ht]
\begin{center}
\includegraphics[width=10cm]{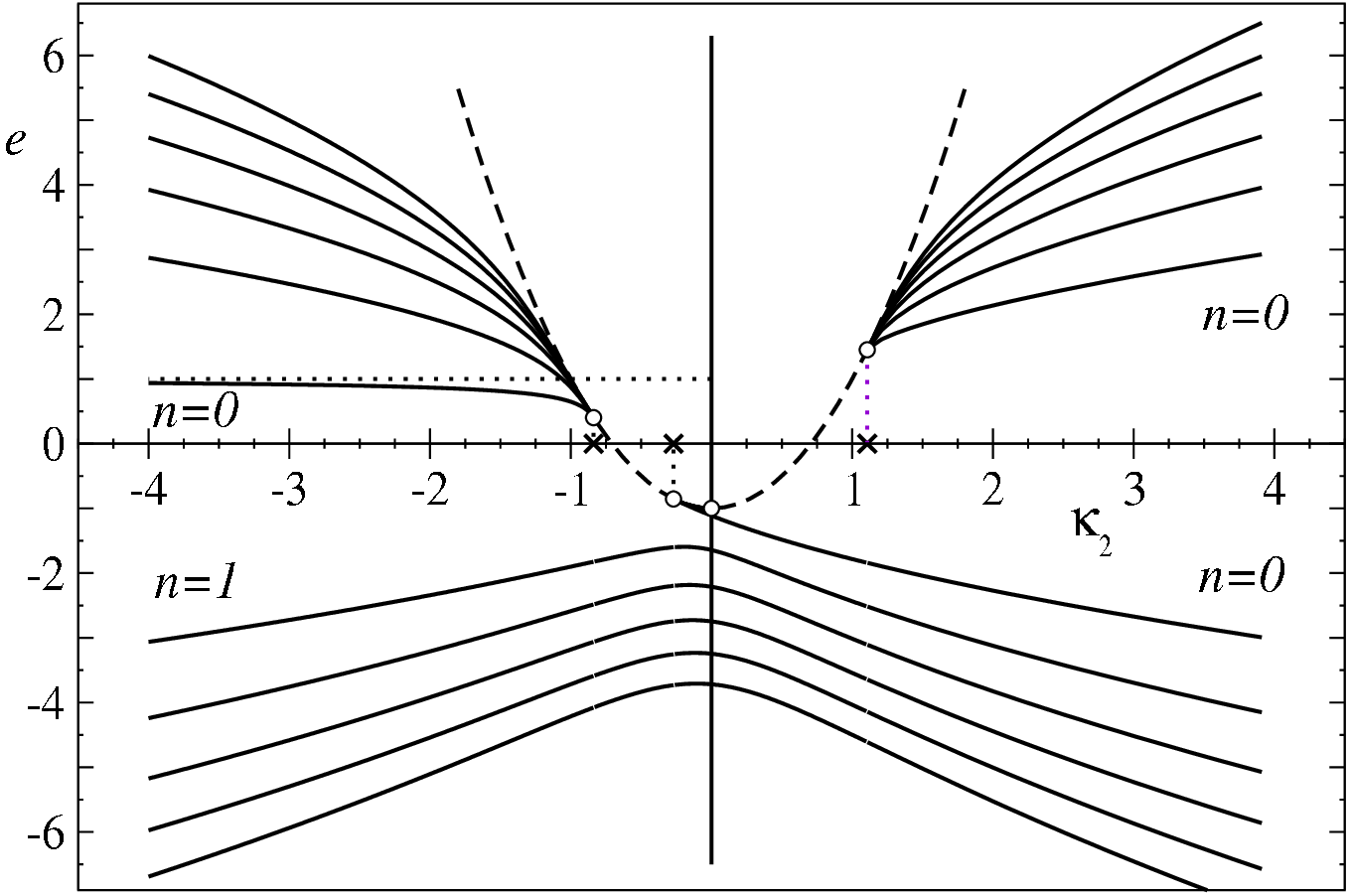}
\end{center}
\caption{Scaled energies for the first six levels when $\protect\kappa_1=-1>\protect%
\kappa_1^c$ as a function of $\protect\kappa_2$. As in Fig.~\protect\ref%
{Fig4} the three non-zero values of the roots of $F(k_2)$, $\protect\kappa%
_2^*$, are marked with a cross in the energy axis. As in that figure, the
hollow circles mark the point ($\protect\kappa_2=0$, $e=-1$) and
the three points where ($k_2=k_2^*$, $e=e_{\rm min}$). The dashed parabola
represents the function $-1-2\protect\kappa_2^2/\protect\kappa_1$ (see text).}
\label{Fig5}
\end{figure}

\begin{figure}[!ht]
\begin{center}
\includegraphics[width=10cm]{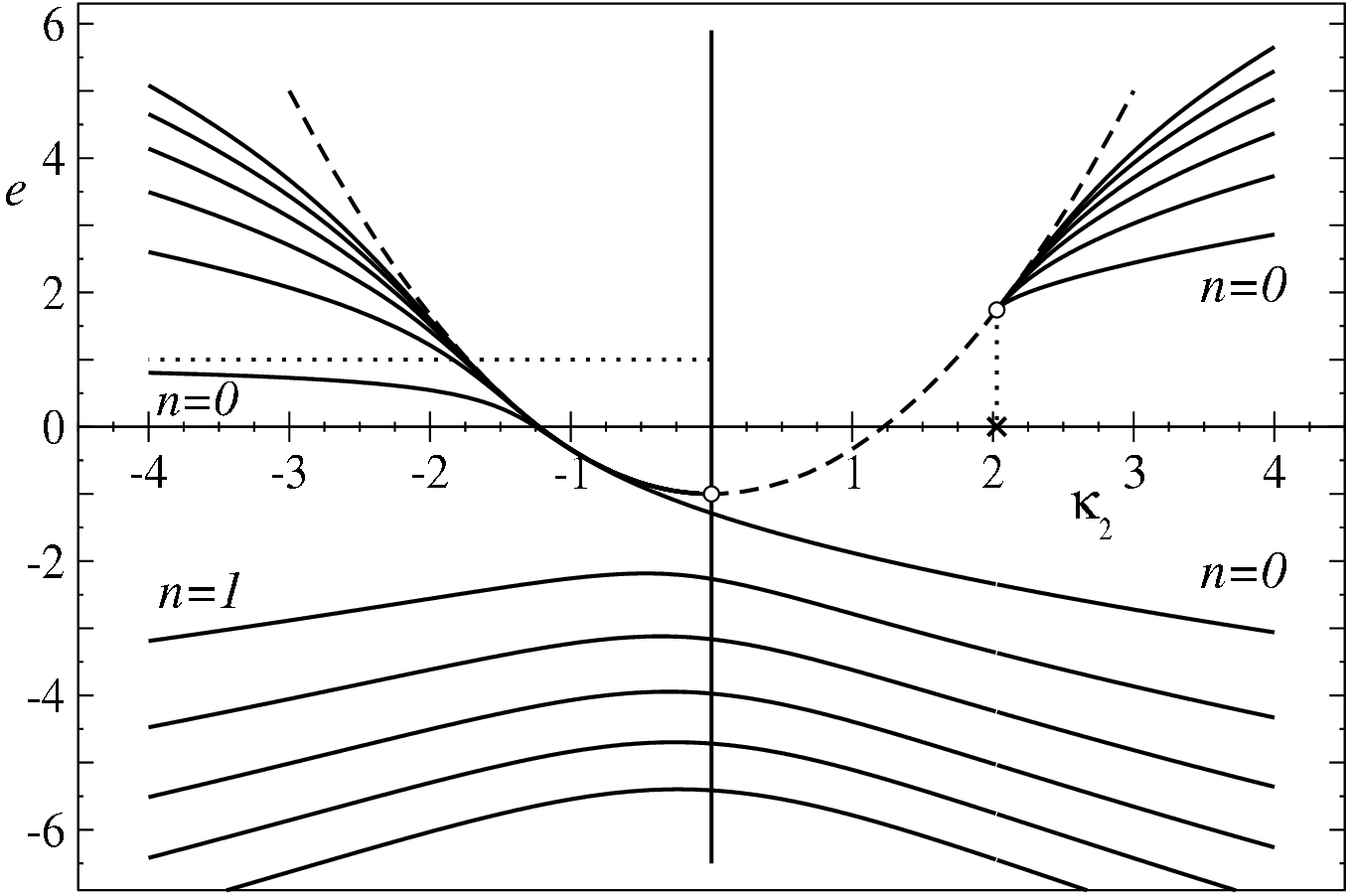}
\end{center}
\caption{Scaled energies for the first six levels when $\protect\kappa_1=-3<\protect%
\kappa_1^c$ as a function of $\protect\kappa_2$. The conventions and meaning of the symbols
are the same as in Figs.~\protect\ref{Fig4} and \ref{Fig5}.
The dashed parabola represents the same function as in Fig.~\ref{Fig5}.}
\label{Fig6}
\end{figure}

In addition to that set of solutions, there can be another set of solutions only if
the function $F(k_{2})$ (see Eq.~\ref{def_F(k2)}) is zero or positive, as
explained before. To determine the values of $k_{2}$ for which this happens,
one has to compute the roots $k_{2}^{\ast }$ of $F(k_{2})=0$. When $k_{1}<0$,
Descartes' rule of signs applied to $F(k_{2})=0$ states that one can
have a positive root and two or zero negative roots (this is done by changing
$z\to -z$ in (\ref{Descartes_rule})), in addition to, of course, the
root $k_{2}^{\ast }=0$. The two negative roots arise for a value of $%
k_{1}=k_{1}^{c}$ such that $F^{\prime }(k_{2}^{\ast })=0$. In scaled units,
this value is $\kappa_{1}^{c}=-64/27$, corresponding to the double negative
root $\kappa_{2}^{\ast }=-8/9$. For $\kappa_{1}<\kappa_{1}^{c}$ there is
only a positive root, while for $\kappa_{1}^{c}\leq \kappa_{1}<0$ there
are two negative roots and a positive root. Because $F(k_{2})$ changes sign
at $k_2=\kappa_{2}^{\ast}$ (except when $\kappa_{2}^{\ast}$ is a double root)
and $F(k_{2})$ is always positive when $|k_{2}|\rightarrow \infty $,
this means
that when $\kappa_{1}<\kappa_{1}^{c}$ there is only one range of values of $k_{2}$
where $F(k_{2})<0$ whereas when $\kappa_{1}>\kappa_{1}^{c}$ there are two ranges of
$k_{2}$ values where this happens. Within these ranges,
only negative-energy solutions exist.
When $\kappa_{1}=\kappa_{1}^{c}$,
the value $\kappa_{2}=\kappa_{2}^{\ast }=-8/9$ corresponds to an infinite
set of degenerate negative-energy solutions with scaled energy $%
e=-1-2(\kappa _{2}^{\ast })^{2}/\kappa _{1}^{c}=-5/3$, which, as mentioned before,
are not bound-state solutions. These features of the
spectrum can be seen in Figs. \ref{Fig5} and \ref{Fig6}.

In Fig.~\ref{Fig5} the solutions with $-mc^2<E<0$ with $n=1,\ldots,5$ are
basically degenerate (they follow closely the parabola $-1-2\kappa_2^2/\kappa_1$)
and the $n=0$ solution is very close to them. Since, as remarked before,
on the parabola the value of $K$ is zero,
these are very weakly bounded solutions. Note that the $n=0$
positive-energy solution with $E<mc^2$ goes to $mc^2$ when $k_2\to-\infty$, thus
becoming the isolated solution of the Dirac oscillator ($k_1=0$). This can
also be seen in Fig.~\ref{Fig6}, where one may also remark that there is a
value of $\kappa_2$ around $-0.9$ for which the solutions for distinct
values of $n$ and $-mc^2<E<0$ are almost degenerate. This point corresponds
to the local minimum of $F(k_2)$, which, for $\kappa_1=-3$, is very close to
the double root $\kappa_2^*=-8/9$. Similarly as in the case of $k_1>0$, the
solutions with $k_2=k_2^*\not=0$ have infinite degeneracy, but they do not correspond
to bounded solutions. Again, for $k_2$ near $k_2^*$ there is a high density
of very delocalized states. The parabola $-1-2\kappa_2^2/\kappa_1$
sets now an upper bound ($n\to\infty$) for the positive-energy states.

The limits $k_{2}\rightarrow 0$ and $|k_{2}|\rightarrow \infty $ correspond,
respectively, to the harmonic oscillator solutions for negative-energy
(antiparticle) states and to the Dirac oscillator. In the first case there
are no bound states for particles (positive-energy states). As $|k_{2}|$
increases, positive bound states start to appear.

\subsection{Variation with $k_1$ for $k_2$ fixed}

To complete the study of solutions of equation (\ref{eigen_eq_Delta=0}) we
present now a plot of its solutions when $k_2$ is kept fixed and let $k_1$
vary. Note that the existence of one or two solutions is governed by the
sign of the function
\begin{equation}  \label{def_G(k1)}
G(k_1)=\frac{4k_2}{k_1^2} \bigg(k_{2}^{3}+mc^{2}k_{1}k_{2}-\frac{\hbar
ck_{1}^{2}}{4}\bigg)\ ,
\end{equation}
identical with $F(k_2)$, except that now $k_2$ acts as a parameter and $k_1$
is the variable. The critical value is now $\kappa_2^c=-1$. In Fig.~\ref%
{Fig7} we present a plot of the solutions for $k_2>0$, when there are two
possible values for $k_1^*$, the roots of $G(k_1)=0$. Apart the factor
$4k_2/k_1^2$, this is just a quadratic equation in $k_1$, so that it is easily shown
that when $\kappa_2\le -1$ it has no real solutions. This means that,
since in this case $G(k_1)\ge 0$, there are always
two sets of solutions. Note that now the points in the energy-$k_1$ plane for which
$K=0$ are on the hyperbole $-1-2k_2^2/k_1$, such that only in the region between the
hyperbole branches, where $K$ is positive, bound states may exist.
\begin{figure}[!ht]
\begin{center}
\includegraphics[width=10cm]{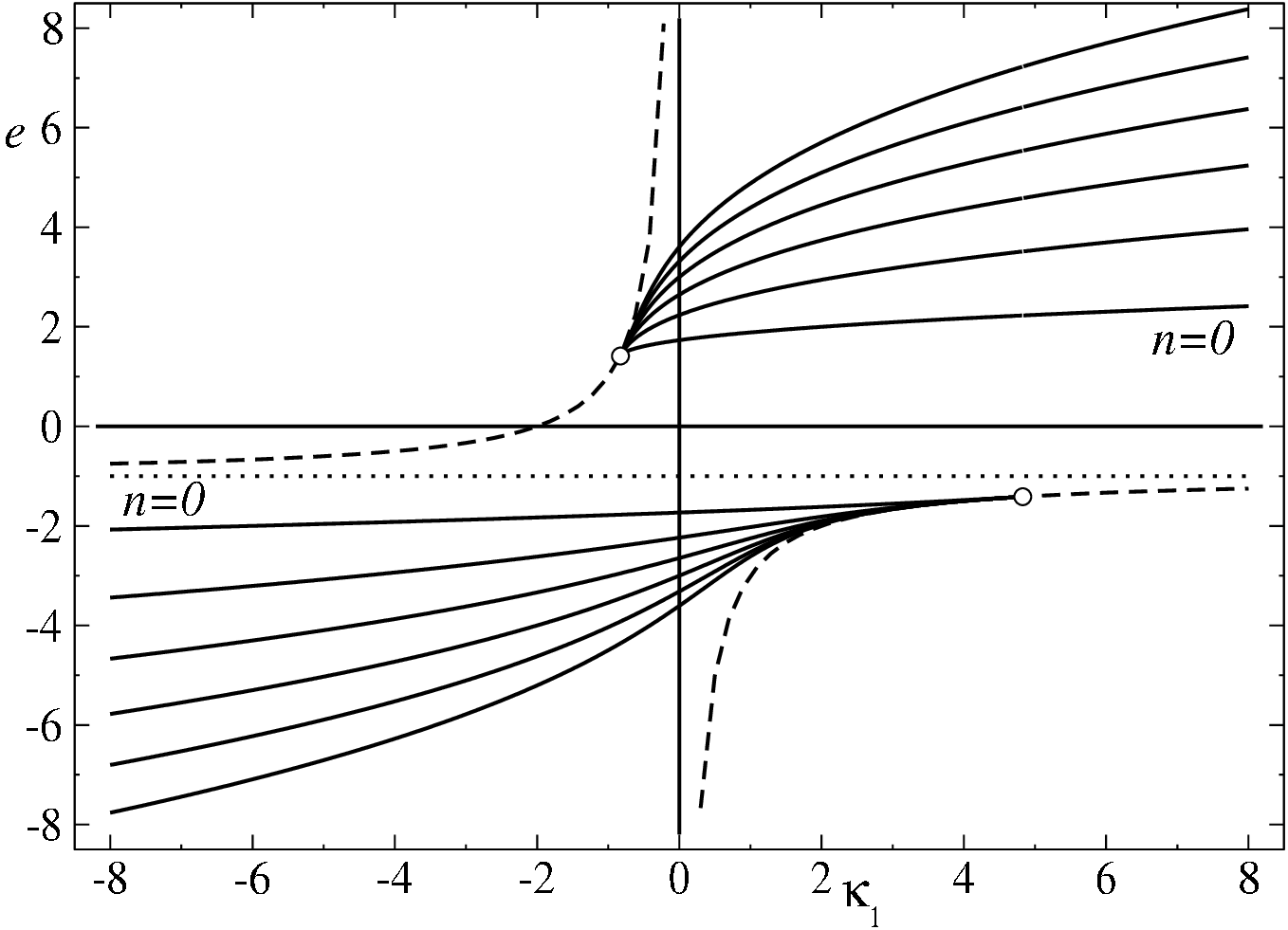}
\end{center}
\caption{Scaled energies for the first six levels when $\protect\kappa_2=1>\protect%
\kappa_2^c$ as a function of $\protect\kappa_1$. As before, the hollow
circles mark the energy solutions for $k_1=k_1^*$. The dashed hyperbole
represents the function $-1-2\protect\kappa_2^2/\protect\kappa_1$ (see text).}
\label{Fig7}
\end{figure}

As before, we can recognize the Dirac oscillator and the harmonic oscillator
limits, as one sets $k_1=0$ and $|k_1|\to\infty$, respectively. In this last
case, one has the harmonic oscillator solutions for particles or
antiparticles depending whether $k_1$ goes to $+\infty$ or $-\infty$. One can
again see that the energy levels of the Dirac oscillator are symmetric about
$E=0$ and that there are no scattering states. Here this follows simply
from the fact that $k_1=0$ is an asymptote of the hyperbole branches which
separate the continuum region from the bound-state region.

\section{Solutions from symmetry arguments}

\label{Sec:sol_sym_arg}

\begin{figure}[!ht]
\begin{center}
\includegraphics[width=10cm]{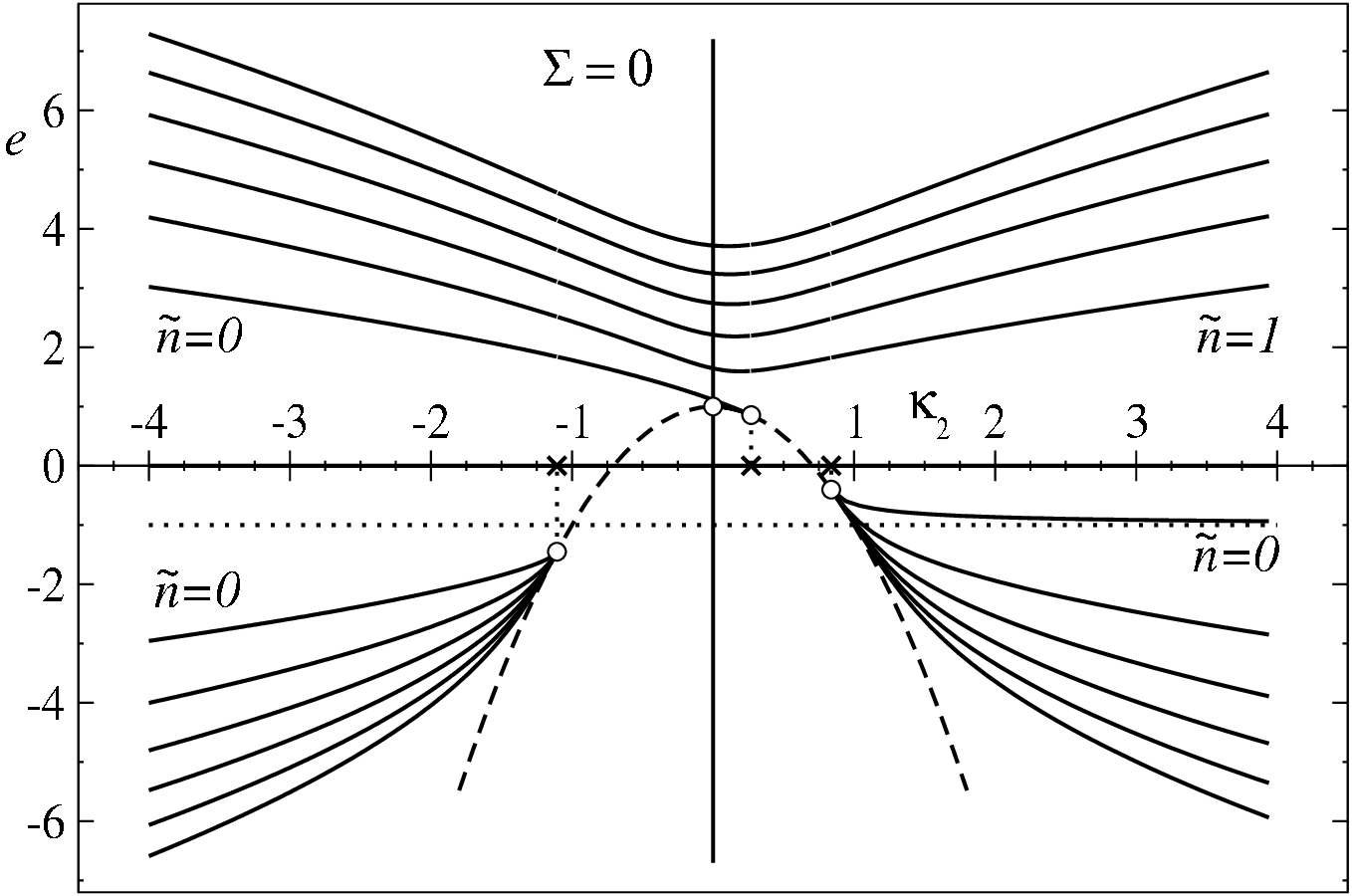}
\end{center}
\caption{Scaled energies for the first six levels when $\protect\kappa_1=1$ as a
function of $\protect\kappa_2$ and $\Sigma=0$. The conventions are the same
as in Fig.~\protect\ref{Fig5}. The dashed parabola represents now the
function $1-2\protect\kappa_2^2/\protect\kappa_1$. $\tilde n$ stands for the
quantum number of the solutions of Eq.~(\protect\ref{eigen_eq_Sigma=0}).}
\label{Fig8}
\end{figure}

As referred before, the case $\Sigma=0$ can be obtained from the $\Delta=0$
case by applying the charge-conjugation transformation. We recall that this
transformation performs the changes $E\to -E$, $\Delta\to -\Sigma$ and $%
V_p\to -V_p$, or, for harmonic oscillator potentials, the changes $k_1\to
-k_1$ and $k_2\to -k_2$. This can be seen if we solve the eigenvalue
equation for $\Sigma=0$, Eq.~(\ref{eigen_eq_Sigma=0}), for $\kappa_1=1$ and
plot the solutions for several values of $\kappa_2$. The result is shown in
Fig.~\ref{Fig8}.

If we compare this figure with Fig.~\ref{Fig5} ($\kappa_1=-1$) we see that
the plots are identical if we reverse both the vertical and horizontal axes,
\textit{i.e}, if we reverse the sign of the energy and of $k_2$,
respectively. Of course, the identification of the particle and antiparticle
levels is also reversed, as it should be, since we are applying the
charge-conjugation transformation.

The existence of both positive- and negative-energy solutions in a system
with $\Sigma=0$ can be relevant for nuclei.
In the 3+1 Dirac equation of relativistic nuclear mean-field theories
there is a connection between the (isoscalar) vector and
tensor potential. The latter is
proportional to the derivative of the former \cite{nosso2}. For the harmonic oscillator
potentials used in this paper, this would give, using the notation of Ref.~%
\cite{nosso2}, the relation
\begin{equation}
\kappa_{2}=\frac{f_{v}}{4}\,\kappa_{1}\ .  \label{k_2_k_1_f_v}
\end{equation}%
The constant $f_{v}$ can take values up to around 1.3 in nuclei using
relativistic mean-field potentials of Woods-Saxon type \cite{nosso2}. For
harmonic oscillator mean-field potentials, there should be also
an upper value for $f_v$, and therefore the relation (\ref{k_2_k_1_f_v})
sets a maximum for $\kappa_{2}$ in nuclei. As was seen above, this is
relevant to know whether there can be simultaneously nucleon and antinucleon bound
states in nuclei. This would depend very much on the strength of the $\Delta
$ potential, which is given by $k_1$. As is apparent from Fig.~\ref{Fig8}, for
small values of the scaled $k_1$, $\kappa_{1}$, the maximum value of $\kappa _{2}$
allowed by Eq.~(\ref{k_2_k_1_f_v}) could be such that there would exist only
positive-energy states. On the other hand, for stronger $\Delta$ potentials,
the relation (\ref{k_2_k_1_f_v}) could allow both positive- and
negative-energy bound states, as can be seen by
applying the axis reversal described above to the $\Delta=0$ spectrum shown in
Fig.~\ref{Fig6}.

\begin{figure}[!ht]
\begin{center}
\includegraphics[width=10cm]{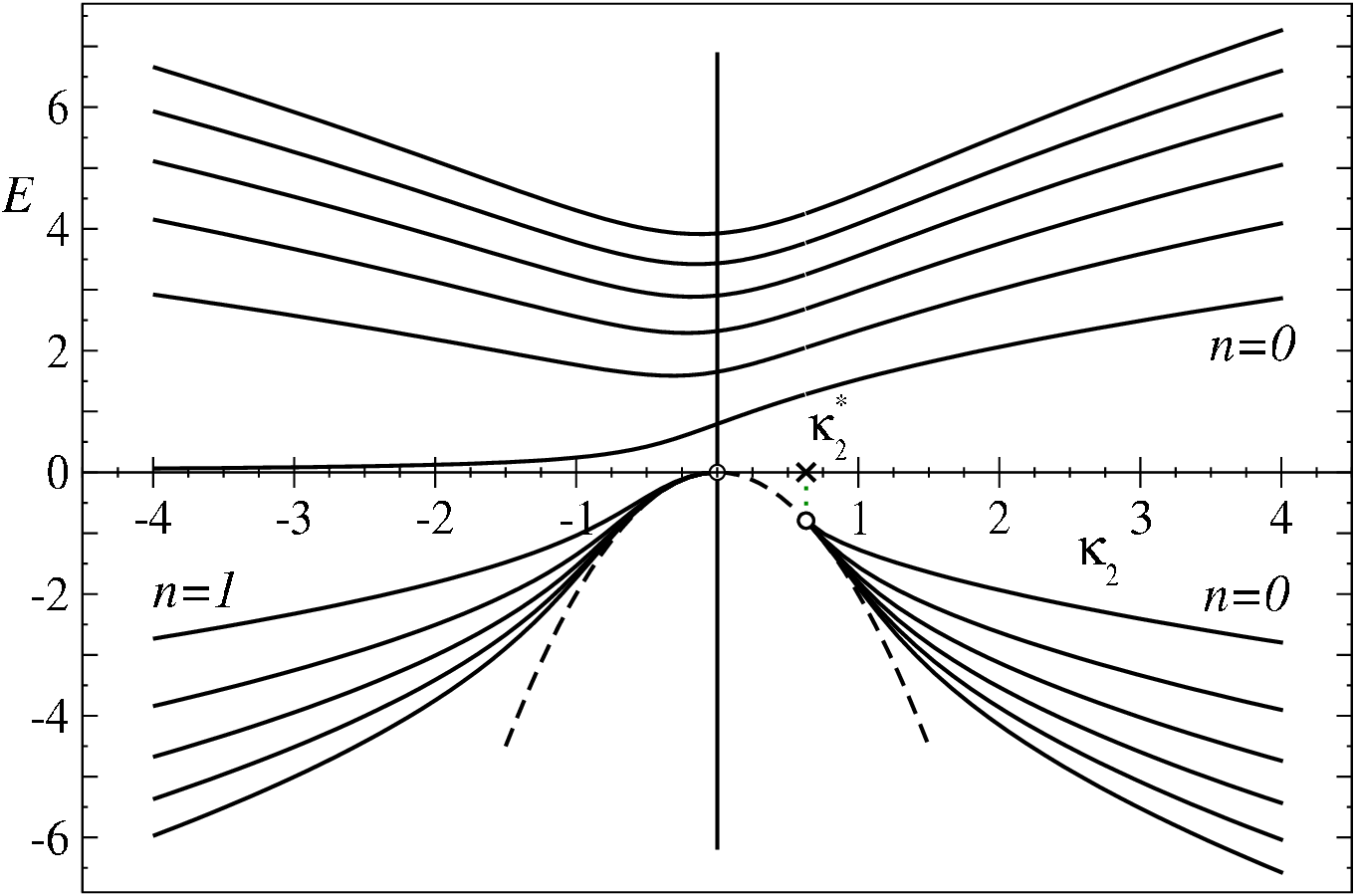}
\end{center}
\caption{Energies for the first six levels when $\protect\kappa_1=1$ as a
function of $\protect\kappa_2$ for $\Delta=0$ and $m=0$. The conventions are
the same as in Fig.~\protect\ref{Fig4}. The dashed parabola represents the
function $-2\protect\kappa_2^2/\protect\kappa_1$.}
\label{Fig9}
\end{figure}

\begin{figure}[!ht]
\begin{center}
\includegraphics[width=10cm]{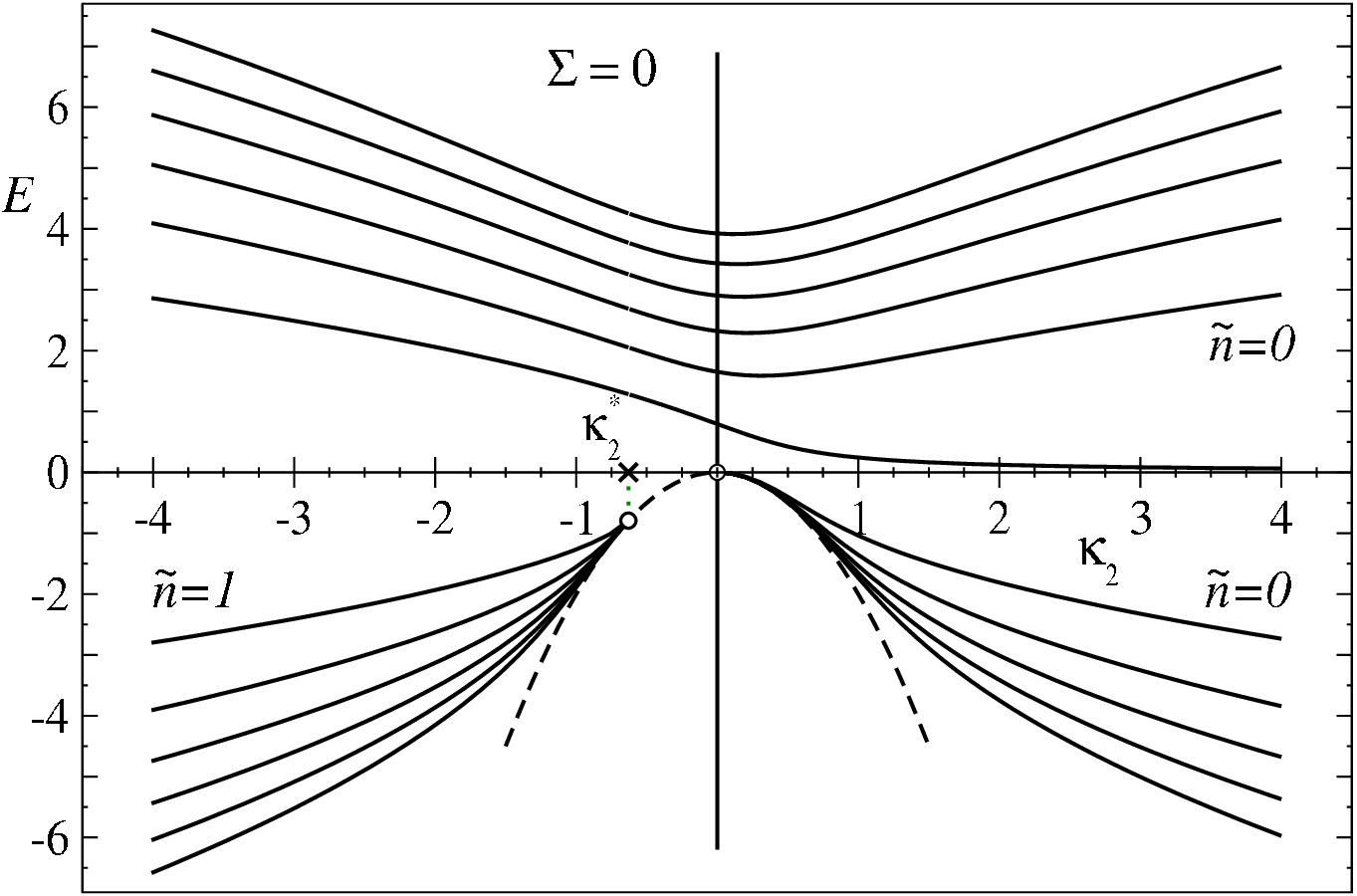}
\end{center}
\caption{Energies for the first six levels when $\protect\kappa_1=1$ as a
function of $\protect\kappa_2$ for $\Sigma=0$ and $m=0$. The conventions are
the same as in Fig.~\protect\ref{Fig4}. The dashed parabola represents the
function $-2\protect\kappa_2^2/\protect\kappa_1$.}
\label{Fig10}
\end{figure}

In Sec.~\ref{Sec:rel_h_oscillator} we have seen that the $\gamma^5$ chiral
transformation can also be used to get formally $\Sigma=0$ from $%
\Delta=0$ solutions, involving, however, the unphysical transformation $m\to
-m$. This objection is overcome when we deal with ultrarelativistic
particles, for which we may consider $m\sim 0$. In this case, the chiral
transformation relates massless physical states with $\Delta=0,\,\Sigma=V$ and a
pseudoscalar potential $V_p$ to massless states with $\Sigma=0,\,\Delta=V$ and
$-V_p$, where $V$ can be an arbitrary potential. For harmonic
oscillator potentials, one has $\Sigma=k_1 x^2/2\,\to\,\Delta=k_1 x^2/2$ and
$k_2\to -k_2$. Since, as explained in Sec.~\ref{Sec:rel_h_oscillator}, the
chiral transformation does not change the equations of motion, the energy of
their solutions is not changed. This can be seen in Figs.~\ref{Fig9} and
\ref{Fig10}, where are plotted, respectively, the solutions of the $\Delta=0$ and $%
\Sigma=0$ eigenvalue equations, Eqs.~(\ref{eigen_eq_Delta=0}) and (\ref%
{eigen_eq_Sigma=0}), when $m=0$, for $\kappa=1$ and several $k_2$ values.
The basic features of these plots were already discussed in the previous
section. As before, the $E=0$ isolated solution was excluded.
We see that the spectra are the same apart from the $k_2\to -k_2$ change,
which shows up as a reversal of the horizontal axis. For $k_2=0$, the pure
harmonic oscillator solution, both spectra are \textit{identical}, with no
negative-energy solutions and, of course, no nonrelativistic limit. Note, however,
that the $\Delta=0$ and $\Sigma=0$ energy eigenstates \textit{are not
the same}, since $\gamma^5$ interchanges the upper and lower components.
This amounts to say that these states are not eigenstates of $\gamma^5$,
which of course they could not be, since, because of $V_{s}$, it does not
commute with the Hamiltonian (\ref{hamiltonian_d_sigma_delta}) even with $%
m=0 $ and $V_p=0$. If $V_s$ would be zero, then, because one has either $%
\Delta=0 $ or $\Sigma=0$, $V_t$ would have also to be zero, and we would
have a free ultrarelativistic particle, which has a good chirality, \textit{%
i.e.}, is an eigenstate of $\gamma^5$.

\section{Conclusions}

\label{Sec:conclusions}

In this work we studied the solutions of 1+1 Dirac equation with a potential
with the most general Lorentz structure which leads to harmonic oscillator
potentials in the second-order differential equation for the spinor components.
This is achieved when the
vector $V_t$ and scalar $V_s$ potentials are quadratic and satisfy either
$\Delta =V_{t}-V_{s}=0$ or $\Sigma =V_{t}+V_{s}=0$, and the pseudoscalar potential is
linear. We showed that there are no bound-state solutions in the
nonrelativistic limit for $\Sigma =0$ and $V_p$ very small, meaning in this case we have an
intrinsic relativistic problem.

We analyzed in great detail, by use of a graphical procedure, the solutions of the
transcendental eigenvalue equation in the case of $\Delta=0$, for all signs of the
potentials and for both particle and antiparticle states. As referred in the
Introduction, there is an increasing interest in studying the behavior of
antinucleon in nuclei, and in particular how well pseudospin and spin
symmetries hold for nucleons and antinucleons respectively.

We discussed in detail the isolated solutions, \textit{i.e.}, solutions out
of the Sturm-Liouville problem, as well as the critical points in the parameter
space of $k_1$ and $k_2$, the coefficients of
the quadratic and linear potentials. Near these points
there is a high density of bounded states which are very extended in space.
These critical points are also relevant to determine which values of
$k_1$ and $k_2$ allow to have a spectrum of both fermion and antifermion bounded solutions
simultaneously or just one of these kind of solutions. We established a connection to 3+1
dimensional relativistic oscillator harmonic potentials applied to nuclei,
where there is a relation between those coefficients and concluded that, for
a sufficiently strong $\Delta$ potential, both kinds of
solutions (\textit{i.e}, for both nucleons and antinucleons) can coexist.

We showed how to obtain the solutions for $\Sigma=0$ from the $\Delta=0$
case, using the charge-conjugation and chiral transformations.
In particular, we used the chiral transformation
to show that, for massless fermions and zero pseudoscalar potential,
the $\Delta =0$ and $\Sigma =0$ solutions have the same spectrum. These conclusions
are quite general and do not depend on the number of dimensions so that they
remain true in 3+1 dimensions. Also, most of the features of the 1+1 spectra
presented in this work are also present in the 3+1 relativistic
harmonic oscillator. Indeed, since the 1+1 eigenvalue equations
for $\Delta=0$ and $\Sigma=0$ are very similar to
the corresponding 3+1 equations, we can apply the same graphical methods
to the 3+1 case, and therefore most of the analysis of the 1+1 case remains true with appropriate
changes. For instance, the shape of the parabola in the energy-$k_2$ plane related to $E_{\min}$
would depend now on the value of the spin-orbit quantum number $\kappa$.
The function $F(k_2)$ would also depend on $\kappa$, and therefore the roots
$k_2^*$ and critical value $k_2^c$ would also depend on $\kappa$. For $k_2=0$ there would
be the known harmonic oscillator $2n+l$ degeneracy, produced by the U(3) symmetry
studied very recently by Ginocchio \cite{gin6} in connection to spin and isospin symmetry.

Finally, the chiral transformation enables us
to switch between the so-called pseudospin ($\Sigma=0$ and $V_p=0$) and
spin symmetries ($\Delta =0$ and $V_p=0$) \cite{bell,gin5} \textit{for any potential}. In the context
of a study of the symmetries of a 3-dimensional relativistic harmonic oscillator, Ginocchio
already showed that one can transform the generators of the SU(2) spin symmetry
into the pseudospin generators by a $\gamma^5$ transformation \cite{gin6}. We showed that
this transformation allows us to relate the $\Delta=0,\,\Sigma=V$ and $\Sigma=0,\,\Delta=V$ problems
for any potential $V$, and, for massless fermions and zero pseudoscalar potential,
also to conclude that the two energy spectra are the same.

\begin{acknowledgments}
We acknowledge financial support from CAPES (PDEE), CNPq, FAPESP, and
FCT (POCTI) scientific program.
\end{acknowledgments}

\end{document}